\documentclass[11pt,a4paper]{article}
\usepackage[makeroom]{cancel}
\usepackage{bm}
\usepackage{amsmath}
\usepackage{amsfonts}
\usepackage{graphicx}
\usepackage{nicefrac}
\usepackage{authblk}
\usepackage{cite}
\usepackage[left=2.2cm,top=2.2cm,right=2.2cm,bottom=2.2cm,nohead]{geometry}
\DeclareGraphicsExtensions{.png,.jpg,.pdf}
\usepackage{epstopdf}
\usepackage{float}
\usepackage{subfigure}
\usepackage{fouriernc} 
\usepackage{enumitem}
\setlist{nolistsep,leftmargin=*}
\usepackage{mdframed}
\usepackage[version=4]{mhchem}
\usepackage{soul}
\usepackage{xcolor}
\definecolor{myblue}{RGB}{212,239,253}

\title{Derivation of an extended Bjerrum equation for the activity coefficient of ions 
based on an analysis of Coulombic forces} 

\usepackage{authblk}
\makeatletter
\renewcommand\AB@authnote[1]{\textsuperscript{\normalfont#1}}

\makeatother

\author{P.M.~Biesheuvel}

\affil{Wetsus, European Centre of Excellence for Sustainable Water Technology, 
The Netherlands}

\date{} 

\begin{document}


\renewcommand{\t}{\widetilde}
\renewcommand{\t}{}
\newcommand{\s}[1]{\mathrm{_{#1}}}

\maketitle

\begin{abstract}
The activity coefficient of ions in solution was proposed by Bjerrum (1916) to depend on the cube root of concentration, because of a good fit with data for low and moderate salt concentration. However, Debye and H\"uckel (DH) later developed a theory that prevailed, and this theory has a square root dependence. The derivation of the DH equation is not easy or intuitive, and has various uncertain elements, and the fit to data is not very good for salts with non-unity valencies. We develop a model for the activity coefficient based on the Coulombic forces between an ion and the nearest ions of opposite charge, including the distribution of separations between these ions. For a symmetric salt, we only have to analyse the interactions between one anion and one cation, and we can derive an expression for the dilute limit that depends on the Bjerrum length and the cube root of salt concentration, the same as an expression put forward by Bjerrum. Our theory also analytically describes higher salt concentrations and a non-zero ion size, not only for 1:1 but also for 2:2 and 3:3 salts. For a 1:1 salt, the cube root scaling law describes data of the activity coefficient very well, both at low and intermediate salt concentrations, while the extended Bjerrum equation describes data even better and up to higher concentrations, and also includes the effect of ion size. For asymmetric salts (2:1 and 3:1 salts), a numerical procedure can be used which considers all possible orbits of two or three monovalent ions around a central multivalent ion of opposite charge, which also fits data accurately. 


\end{abstract}

An important topic in physical chemistry is the activity coefficient of ions in an electrolyte solution. The activity coefficient is the correction to the concentration required to obtain a good prediction of the chemical potential of an ion, $\mu_i$, and other thermodynamic properties of an electrolyte solution, such as the osmotic pressure and vapour pressure. The activity coefficient $\gamma_i$ relates to activity $a_i$ and concentration $c_i$ by $a_i=\gamma_i c_i / c\s{ref}$, and for the dimensionless chemical potential we obtain
\begin{equation}
\mu_i = \mu\s{ref,\textit{i}} + \ln a_i = 
\mu\s{ref,\textit{i}} + \ln \left( c_i / c\s{ref} \right) + \ln \gamma_i \, .
\end{equation}
All contributions to $\mu_i$ can be multiplied by $RT$ to arrive at a potential expressed in J/mol. 

Typical results for $\gamma_i$ are that starting at a value of $\gamma_i \!= \! 1$ for a very dilute solution it first decreases with salt concentration and then increases again. We can also evaluate the term $\ln\gamma_i$, which starts at zero, becomes more and more negative with increasing salt concentration, and then increases again to positive values beyond a certain salt concentration. We discuss both the initial decrease which has an electrostatic origin, and the subsequent increase which is due to volumetric interactions. 
The electrostatic contribution is because anions and cations attract one another by Coulombic forces, and the more so at higher concentrations, 
and this lowers the electrostatic energy and leads to a contribution to the chemical potential of an ion. 

\begin{figure}
\centering
\includegraphics[width=0.85\textwidth]{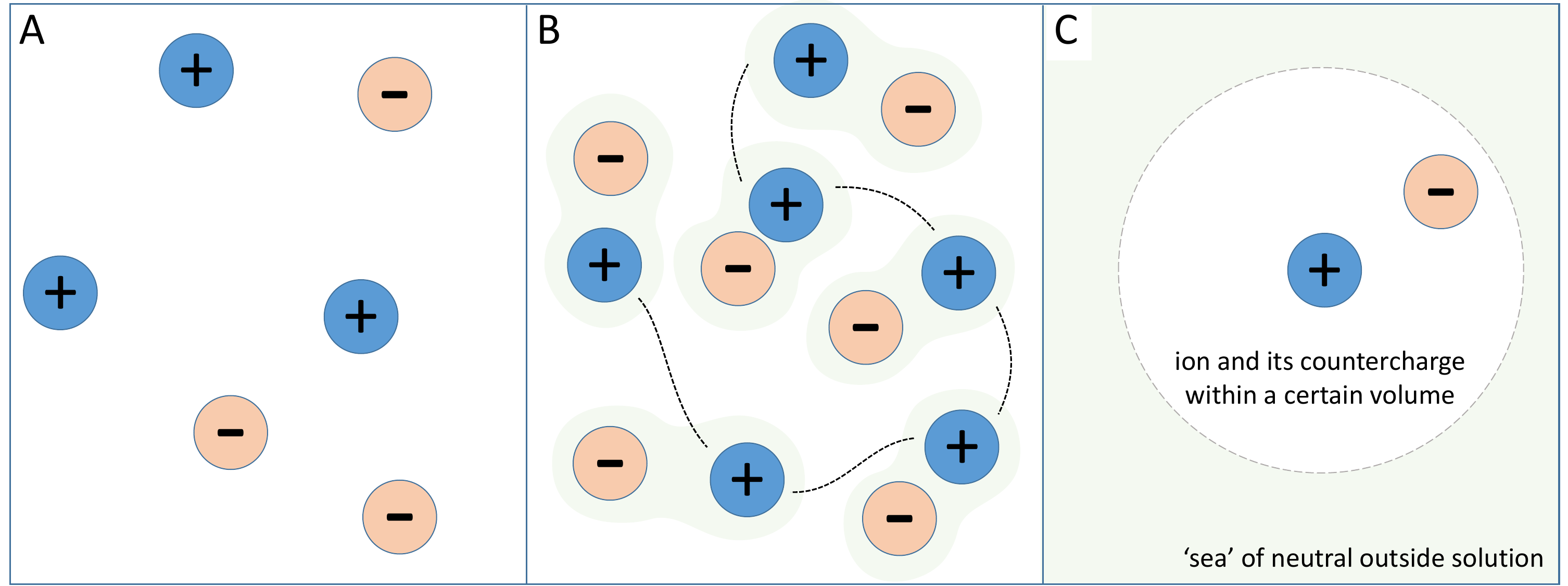}
\vspace{-8pt}
\caption{In a dilute electrolyte solution, ions are randomly distributed as depicted in panel A, but the electrostatic energy is lowered at higher concentrations because ions of opposite charge reside near one another increasingly often, see panel B. In a different interpretation, see panel C, the countercharge to a certain ion (the nearest ion of opposite charge sign) resides within a certain volume, and on the outside of that volume, the charge-neutral solution affects both ions equally. The volume in which the first countercharge is found, decreases when the salt concentration goes up. }
\label{fig_soft_ion_pairs}
\end{figure}

For very large dilutions, all ions are distributed randomly over space without correlations between their trajectories, as depicted in Fig.~\ref{fig_soft_ion_pairs}A. But if we increase the salt concentration, ions are on average nearer to one another, and then it becomes more likely for them to seek out positions even nearer to ions of opposite charge, and this reduces the electrostatic energy, see Fig.~\ref{fig_soft_ion_pairs}B. The assumption now is that to calculate the electrostatic energy of the system, we only have to consider an ion and its nearest countercharge. That first countercharge resides somewhere in a volume around the center ion. That volume defines where the first counterion is found, and this volume will go down at higher salt concentrations, because it becomes more likely that the first counterion is closer. All other ions are outside this volume, and together are charge neutral and their motions are not correlated to those in the central volume, so they do not influence in any way the interactions between these key ions. So the two or more central ions (for a 1:1 salt one anion and one cation) are associated to one another, and we can call this an ion pair or ion ensemble. But this association is not of a chemical kind, but is solely due to a balance between Coulomb's law that is attractive between anions and cations, and random motion,  
which tends to distribute all ions over space. Which ion is another ion's countercharge will also 
rapidly change over time. The association of two ions in a pair is merely `statistical', defined by being for some time one's nearest countercharge. 

Thus, to calculate the electrostatic energy of a system, we 
only have to solve the Coulombic forces between one ion and its most nearby countercharges. For instance for a 2:1 salt this means we only consider the interaction of one divalent cation and two monovalent anions. We assume we can neglect all Coulombic interactions with the `outside' ions that are further away. This must be a correct assumption when the attractions and repulsions to these outside ions exactly cancel when averaged over time. 

We calculate the energy decrease due to Coulombic interactions in such an ion pair or ion ensemble. We first discuss symmetric $z \! : \! z$ salts, in which case we only have to consider a pair of ions, 
such as depicted in Fig.~\ref{fig_soft_ion_pairs}C. Later on we discuss asymmetric 2:1 and 3:1 salts and the related formation of ion \textit{triplets} and ion \textit{quadruplets}. When describing an ion pair, we must know what are the possible ion-ion separations, \textit{x}, and the likelihood of each separation. The anion and cation in the ion pair cannot be closer than the sum of their radii, so that is a minimum separation. The maximum separation is harder to estimate, but the countercharge to an ion cannot be further away than a distance, $x\s{max}$, at which the `sea' of outside ions start, see Fig.~\ref{fig_soft_ion_pairs}C. 
%
This range of possible separations, up to $x\s{max}$, is compressed when the salt concentration increases, because the higher the concentration, the closer on average is the second nearest ion of opposite charge to be found. So with increasing salt concentration the sea of outside ions is `closing in'. 
It is only in a statistical sense that the ions in a pair are pushed together, into  a smaller volume, when the salt concentration increases. 
This approach in which salt concentration $c_\infty$ determines the volume in which an ion and its countercharge are found, is of key importance in our theory. We have not seen this concept before in studies of Coulombic interactions in electrolyte solutions. 

Based on this approach, we can calculate the ion pair energy, IPE. This energy we can multiply with $c_\infty$ (because that is how many ion pairs there are per unit volume) to obtain the free energy density relating to this Coulombic effect, and we take the derivative with $c_\infty$ and divide by 2, to find the contribution of direct ion-ion Coulombic forces to the mean activity coefficient of an ion.

If we integrate the force between two opposite charges according to Coulomb's law, from infinity to a separation \textit{x}, for a symmetric $z \! : \! z$ salt, we find that the electrostatic energy of the ion pair (in units~{kT}) is 
\begin{equation}
\psi(x) = - z^2 \cdot \lambda\s{B} \, / \, {x}
\label{eq_corr_2}
\end{equation}
where the Bjerrum length is $\lambda\s{B}={e^2}/{4\pi \varepsilon\s{r} \varepsilon_0 kT}$. At $T\!= \! 298$~K, the dielectric constant of water is $\varepsilon\s{r}\!=\! 78.3$ and thus $\lambda\s{B}\!= \! 0.716$ nm.  

The probability distribution function \textit{f}(\textit{x}) of finding the two ions at a separation \textit{x} is
\begin{equation}
f(x) = \frac{\phantom{\int}\exp\left(-\psi(x)\right) \cdot x^2 \; \phantom{\text{d}x}}{\int \exp\left(-\psi(x)\right) \cdot x^2 \; \text{d} x}  
\label{eq_corr_3}
\end{equation}
and combination with Eq.~\eqref{eq_corr_2} shows that the larger is \textit{z}, or the larger is $\lambda\s{B}$ (because the temperature or the dielectric constant is lower), the more likely it is that ions are close to one another. This equation was derived by Bjerrum (1926), 
but Bjerrum evaluates it by integrating to a maximum distance that depends on the Bjerrum length, and he then proceeds with a model based on ion association equilibria. Instead, in our work the upper integration limit is \textit{x}\textsubscript{max}, which depends on $c_\infty$. When $c_\infty$ goes up, this maximum separation \textit{x}\textsubscript{max} goes down and this effectively pushes ions together, and thus IPE becomes more negative. The minimum separation is equal to the sum of the two ion radii, $x\s{min}=a_+ + a_- = 2 \left\langle a \right\rangle$. In this distribution function, in the absence of the Coulombic force, all positions in the space between \textit{x}\textsubscript{min} and \textit{x}\textsubscript{max} are equally likely (per volume element). The \textit{x}\textsuperscript{2}-term is because of the spherical geometry, which also has the effect that two ions being very close is unlikely.  

The ion pair energy, IPE, is given by 
\begin{equation}
\text{IPE} = \int f(x) \psi(x) \; \text{d}x
\label{eq_corr_4}
\end{equation}
and after inserting Eqs.~\eqref{eq_corr_2} and~\eqref{eq_corr_3}, we obtain
\begin{equation}
\text{IPE} = - \frac{\int \exp\left(\beta/x\right) \cdot x  \cdot \beta \; \text{d}x }{\int \exp\left(\beta /x\right) \cdot x^2 \hphantom{\cdot \beta \;} \text{d} x} 
\label{eq_ipe_full}
\end{equation}
where $\beta=z^2 \lambda\s{B}$. 
Series expansion of Eq.~\eqref{eq_ipe_full} around $1/x\s{max}=0$ results in 
\begin{equation}
\text{IPE}= - \frac{\int \left( \beta \cdot x + \beta^2 + \nicefrac{1}{2} \beta^3 / x + \nicefrac{1}{6} \beta^4 / x^2 + \dots \right) \; \text{d}x }{\int \left(  x^2 + \beta \cdot x + \nicefrac{1}{2} \beta^2 + \nicefrac{1}{6} \beta^3 / x + \dots \right) \; \text{d} x} 
\label{eq_corr_4a}
\end{equation}
and if we then only take the first term in each series, and assume $x\s{max}\gg a$, we obtain after integration
\begin{equation}
\text{IPE} 
= - 3/2 \cdot z^2 \cdot \lambda\s{B} \, / \, x\s{max} \, . 
\label{eq_corr_5}
\end{equation}

Now we must discuss how the maximum separation \textit{x}\textsubscript{max} relates to salt concentration $c_\infty$ (in mM = mol/m\textsuperscript{3}) or $n_\infty$ (unit m\textsuperscript{-3}), with $n_\infty  = c_\infty \, N\s{av}$. When solutes do not interact, the average distance between them scales with $n_\infty^{-\nicefrac{1}{3}}$, i.e., an 8$\times$ increased concentration reduces the average distance by a factor of 2. Therefore also $x_\text{max}$ can be expected to scale with $n_\infty^{-1/3}$. This proportionality can be rewritten to $n_\infty  = 1/\left(\alpha \cdot x\s{max} \right)^{3}$, with $\alpha$ a yet to be defined factor. 
To find $\alpha$, we evaluated many geometries over the years of how ions are spaced and how to calculate the probability distribution of finding the first and second counterions, and some of these geometries result in values of $\alpha$ close to unity, but the assumptions on which these geometries were based were never very convincing. But approaching the problem empirically, we find that for symmetric salts, $\alpha \! = \! 1$ results in an excellent fit of the theory to the data, and it is therefore that we will use this value of $\alpha$ for calculations involving $z \! : \! z$ salts. (And as we discuss further on, to make a generalization to also include asymmetric salts, in the equation for $x\s{max}$ we replace $c_\infty$ by the total ions concentration divided by two.) 

We continue with a $z\!:\!z$ salt and implement this relation between $n_\infty$ and \textit{x}\textsubscript{max} in Eq.~\eqref{eq_corr_5}, i.e., $n_\infty = x_\text{max}^{-3}$, multiply IPE by $n_\infty$ (because $n_\infty$ is the concentration of ion pairs) to obtain the free energy density \textit{f}, of which we take the derivative with $n_\infty$, and then divide by 2, which leads to the contribution of this Coulombic effect to each ion's chemical potential. We then arrive at
\begin{equation}
\ln\gamma_\pm = \mu\s{i.i.c.i.} = -  \alpha \cdot z^2 \cdot \lambda\s{B} \cdot \sqrt[3]{n_\infty}
\label{eq_ipe}
\end{equation}
where $\ln \gamma_\pm$ is the mean activity coefficient, and we use index i.i.c.i. for `ion-ion Coulombic interactions', see ref.~\cite{Biesheuvel_2022}. A simplified notation of Eq.~\eqref{eq_ipe} is
\begin{equation}
\ln\gamma_\pm = - b \cdot {c_{\infty}}^{1/3} 
\label{eq_mu_11_practical}
\end{equation}
where the factor $b$ is given by $b = \alpha z^2 \lambda\s{B} N_\text{av}^{1/3}$ with $N\s{av}$ Avogadro's number, $N\s{av}=6.022 \cdot 10^{23}\text{~mol}^{-1}$. This factor \textit{b} depends on temperature and valency of the symmetric salt. For a 1:1~salt at $T\!=\!298$~K, we have $b \! = \! 0.0605\text{~}\left(\text{mM}\right)^{-1/3}$. As we will discuss, for a 1:1 salt, 
Eq.~\eqref{eq_mu_11_practical} accurately describes data for the activity correction, from the dilute limit up to quite concentrated solutions of $c_\infty \! = \! 200$~mM, and in this entire range $\ln \gamma_\pm$ therefore scales with the cube root of salt concentration. As we will also discuss, for asymmetric \textit{z}:1 salts, this cube root law works equally well, but the prefactor $b$ is larger, and its value must be derived from the full numerical model. For symmetric $z \! : \! z$ salts with $z \! > \! 1$, the situation is very different, and though Eq.~\eqref{eq_mu_11_practical} is theoretically still valid in the dilute limit, it does not apply in the range of salt concentrations where data are available. Instead, for these $z \! : \! z$~salts we must numerically solve Eq.~\eqref{eq_ipe_full}, or evaluate many terms in Eq.~\eqref{eq_corr_4a}. In Appendix~I we present details of the analytical method to calculate $\ln\gamma_\pm$ accurately for all $z\!:\!z$ salts. 

Eq.~\eqref{eq_ipe} is 
identical to an equation in Bjerrum (1916, 1919) with the same dependence on \textit{z}\textsuperscript{2}, on dielectric constant, and on salt concentration, and with the same prefactor (like in our case, derived empirically),\footnote{This is not the Bjerrum theory of 1926 and thereafter that describes activity coefficients using ion association equilibria.} see also p.~273 in Dole (1935). 
We can also derive an expression for the osmotic coefficient $\phi$ based on this theory. The osmotic coefficient is the real osmotic pressure divided by the ideal osmotic pressure, thus $\phi=\Pi/\Pi\s{id}$. For a symmetric salt, the ideal osmotic pressure is $\Pi\s{id}=2c_\infty RT$. We can use the Gibbs-Duhem equation, $\partial \Pi / \partial c_\infty = 2\, c_\infty \cdot \partial \mu_i / \partial c_\infty$ (for a $z \! : \! z$ salt), and based on Eq.~\eqref{eq_mu_11_practical} obtain
\begin{equation}
\phi= 1 + \frac{1}{c_\infty} \int_0^{c_\infty}  \frac{\partial \ln \gamma  \hspace{7pt}  }{\partial \ln c_\infty} \text{d}c_\infty = 1 - \nicefrac{1}{4} \cdot b \cdot \sqrt[3]{c_\infty} 
\label{eq_osm_pressure_cell_model_11}
\end{equation}
and also this result can be found in Bjerrum (1916, 1919). Eq.~\eqref{eq_osm_pressure_cell_model_11} implies that for a salt concentration in M, we obtain for a 1:1 salt $\phi = 1 - 0.151 \cdot \sqrt[3]{c_{\infty,\text{M}}}$. 
Experimental values for this factor 0.151 were calculated by Bjerrum (1919) based on measurement of freezing point depression by Noyes and Falk (1910), and vary `between 0.146 and 0.225 the mean being 0.17'. This range of values is in very good agreement with our theoretical value of 0.151.\footnote{We can recalculate what temperatures must have been measured. The equation to use is $\Delta T = 2 \, c_\infty \, K\s{F} \, / \, \phi $, where the cryoscopic constant of water is $K\s{F}=1.86$~$^\circ$C/M. Thus, for an experiment at $c_\infty=50$~mM, the temperature decrease was $\Delta T= 0.176 $~$^\circ$C, which in the ideal case would have been $\Delta T\s{id}=0.186$~$^\circ$C, thus a difference of 10~mK. For an experiment at 100~mM, the difference is much larger, with the measured value $\Delta T=0.346$~$^\circ$C and an ideal value of $\Delta T\s{id} = 0.372 $~$^\circ$C, a difference of 26~mK. Thus, very tiny temperature differences were measured to derive values of the osmotic coefficient, which led Bjerrum to propose his cube root law dependence of $\ln \gamma_\pm$ on salt concentration.} A more detailed analysis of osmotic pressure according to the Bjerrum theory and an extended version, and comparison with data for KCl and NaCl, is presented in Appendix~IV.

The dependence of $\ln\gamma_\pm$ on salt concentration to a 1/3$^\text{rd}$ power was proposed and discussed in the early 20\textsuperscript{th} century by Bjerrum (1916, 1919) and Ghosh (1918), see \mbox{Kort\"{u}m} (1965, p.~175), Bockris and Reddy (1970, p.~269), Erdey-Grúz (1974, pp.~415--424) and Kuznetsova (2002). Earlier still, Milner (1912, 1913) developed a model for the Coulombic interactions in a 1:1 salt along lines similar to described in the present work, including an analysis of the distance distribution of the most nearby ions around a key center ion. His papers describe an energy \textit{h} that has a cube root dependence on salt concentration, which is multiplied by a function \textit{f}, and that product, \textit{hf}, is a linear function of an osmotic coefficient. The function \textit{f} is close to a square root power of \textit{h}, and thus the product, \textit{hf}, scales with the square root of concentration. Bjerrum (1916) in his Fig.~1 gives data and calculations of the osmotic coefficient as function of concentration. One theoretical curve which does not run well through the data is `according to Milner', and in line with our analysis above, we find that this curve indeed follows a square root dependence on concentration. 

Erdey-Grúz (1974) describes how not just at low but also at moderate salt concentrations this cube root dependence has a better agreement with experiments than the square root dependence of Milner (1912, 1913) that in 1923 was also put forward by Debye and H\"{u}ckel (DH). The DH equation is based on solving a linearised version of the Poisson-Boltzmann (PB) equation around a certain key ion, which requires the electrical potential to be less than $\sim\!25$ mV. However, these low potentials are not likely, certainly not in the dilute limit. We solve the Laplace equation in a spherical geometry around an ion, and at high dilutions 
we have for the potential at an ion's surface $ kT / e \cdot z\cdot \lambda\s{B}\,  / \, {a}$. For a monovalent ion with a (hydrated) radius between $a \! = \!0.2$ and $a \! = \!0.3$~nm, this potential is between 60 and 90 mV. For divalent and trivalent salts, this factor is two times and three times higher (but ion radius also larger). Calculations that implemented Booth's equation to describe dielectric saturation, which makes $\varepsilon\s{r}$ decrease close to the ion because of electrical field strengths beyond 0.1~GV/m, these calculations give even dramatically higher potentials. Thus, in the direct vicinity of an ion, because of the high potentials, a linearisation, for instance of the PB equation, will lead to an error. In addition, 
Debye and Hückel calculate the energy of an ion with its diffuse layer around it, and equate it to a contribution to the chemical potential. However, this is not correct, 
but an energy expression must be differentiated over salt concentration to obtain an expression for $\ln\gamma_\pm$ (Kontogeorgis \textit{et al.}, 2018).  
In any case, the DH approach seems highly contrived, to single out a certain ion and then use a model that smears out the ion distribution around it, even though all the countercharge \textit{is just one other ion} (for a symmetric salt), i.e., in reality the countercharge is highly discrete and localized. In the DH model, care is also required to avoid double 
counting because in the calculation an ion is on the one hand the key ion with a finite size, and on the other hand is a point charge in the diffuse layers around all other ions. 
A detailed criticism of the Debye-Hückel theory is provided by Wright (1988, 2007) who in 1988 writes `... how naive the Debye-Hückel primitive model is. ... the complexities of fitting this very simple model into a mathematical framework, and the even greater complexities of solving the mathematics, we can understand how badly stuck we are with a model which is so physically unrealistic.' And in 2007 she concludes that the DH theory is `painfully naïve.' 

Earlier, Frank and Thompson (1959) already present a significant criticism of the DH-equation, arguing that it is certainly invalid at salt concentrations above 10 mM, and that already at 1 mM `the theory has failed,' 
(This concentration as an upper limit of validity of the DH-approach is also stated by Bockris and Reddy (1970, p. 269).)  
Frank and Thompson even conclude that for concentrations as low as 0.1~mM the DH-theory `may be in some trouble,' and `should be expected to fail.' 
Thus they argue that the DH equation is only valid at `extreme dilutions,' and `that there is reason to expect that the Debye-Hückel ion cloud to break down at concentrations much lower than has been imagined,' while `above the concentration of this breakdown, cube-root laws are generally and accurately obeyed by real solutions.' And in another summary of their theoretical analysis of the DH-theory: `An argument may be developed at length to show that the Debye-Hückel theory should be expected to break down when the solution becomes too concentrated, and that ``too concentrated'' for this purpose means, in an aqueous solution of a 1-1 strong electrolyte at room temperature, more concentrated than about 0.001 mole liter\textsuperscript{-1}.' 
If they are right, then when combined with our position that the cube root law must hold in the limit of $c_\infty \rightarrow 0$, not much room is left for a range where a square root dependence 
is perhaps better. 
More than 20 years earlier, Van Laar (1936) describes how precise data based on vapour pressure decrease, 
are much better described by a cube root law than a square root law, and he discusses at length the theoretical problems he identifies in the DH derivation. He concludes (translated from Dutch) `.. only a cube root law remains [a viable option], while the square-root law of Debije can be safely discarded on theoretical and experimental grounds.' In 1966, also Poirier comprehensively summarized all theoretical problems with the DH approach.

\begin{figure}
\centering
\includegraphics[width=1\textwidth]{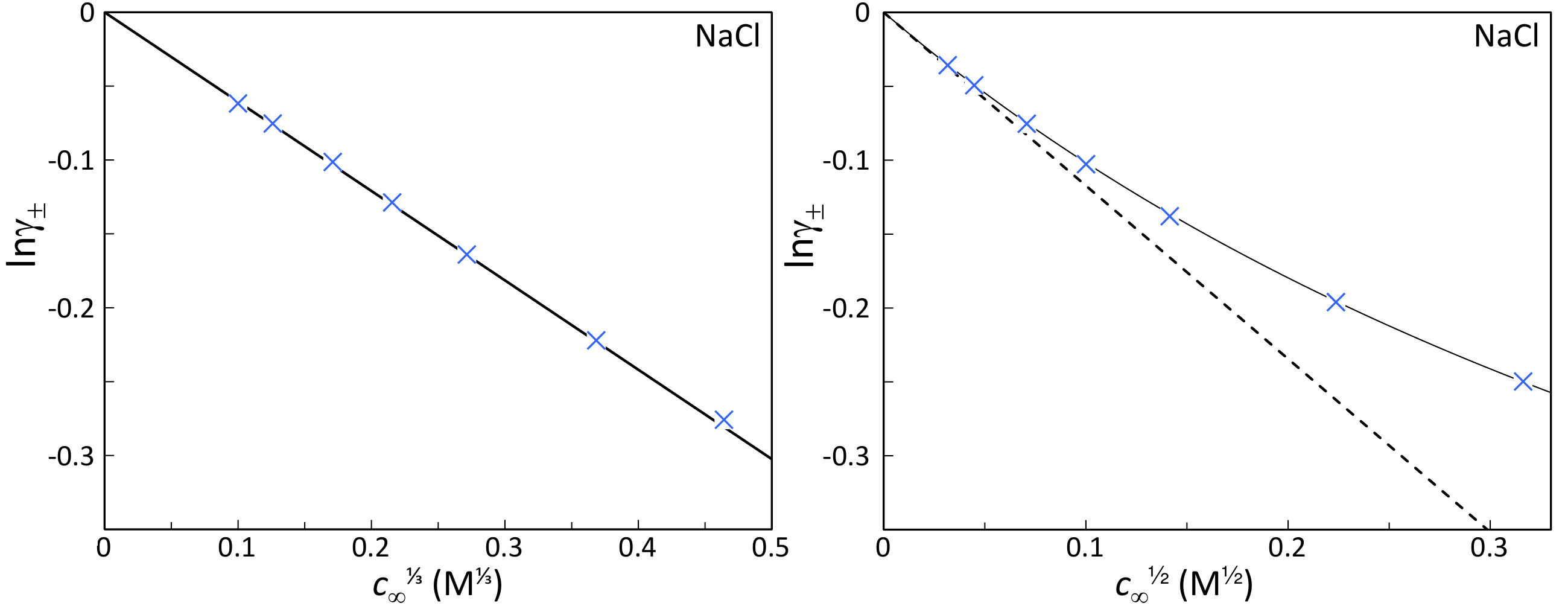}
\vspace{-20pt}
\caption{Data for the ion activity coefficient of NaCl as function of the cube root of salt concentration in panel A and as function of the square root of concentration in panel B. The line in panel A is the Bjerrum equation, Eq.~\eqref{eq_mu_11_practical}. In panel B the dashed line is the Debye-Hückel (DH) limiting law, and the solid line is the extended DH theory with $a_\text{DH}\!=\!0.46$ nm as an effective ion diameter.}
\label{fig_lngamma_NaCl}
\end{figure}

Next we analyse 
the correspondence with experiment of our approach based on calculating the Coulombic energy between nearby ions, and of the DH theory. For NaCl, data of $\ln\gamma_\pm$ vs. $c_\infty$ are presented as Table 16 in Hamer and Wu (1972) for $c_\infty$ between 1 and 100 mM. 
We reproduce these data in Fig.~\ref{fig_lngamma_NaCl} as function of $\sqrt[3]{c_\infty}$ in panel A, and as function of $\sqrt{c_\infty}$ in panel B.  
Panel A shows that $\ln\gamma$ is well described by a linear dependence on $\sqrt[3]{c_\infty}$. The solid line is according to Eq.~\eqref{eq_mu_11_practical} without any further fitting parameter. The published data points were shifted downward, all by the same amount of 0.026~{kT}, retaining how they are positioned relative to one another, which we do in order to make the data extrapolate to $\ln \gamma_\pm \! = \! 0$ when $c_\infty\rightarrow 0$, i.e., to the origin. We discuss this topic further on. The slope of our theoretical line in Fig.~\ref{fig_lngamma_NaCl}A follows from analysis of Eq.~\eqref{eq_mu_11_practical} resulting in  $\partial \ln \gamma / \partial \sqrt[3]{c_{\infty,\text{M}}} = - 10 \cdot b$, 
and this factor is $\sim \! -0.605$~M$^{-1/3}$. A theoretical derivative $\partial \,^{10}\!\log \gamma / \partial \sqrt[3]{c_{\infty,\text{M}}}$ is then $\sim \! -0.26\text{~M}^{-1/3}$, very close to the value of $-0.25$ given by Bjerrum (1916, 1919), see also Glueckauf (1959). Glueckauf points out that experimentally this derivative is `mostly about 0.25, but they vary between 0.225 and 0.305.' Based on the assumption of a rigid ionic lattice, Glueckauf himself comes with a theoretical derivative of $-0.385$, which is too high, and he writes how the greater disorder of ions in solution is likely the cause that the real value is lower than his theoretical prediction. 
Kortüm gives the proportionality factor as $-0.29$, which is 10\% above our theoretical value. Thus, the predicted theoretical slope of $\partial \ln\gamma / \partial \sqrt[3]{c_{\infty}}$ matches experimental observations, in a significant range of concentrations. 
For the Debye-Hückel equation, analysed in Fig.~\ref{fig_lngamma_NaCl}B, the dashed line is the DH limiting law, $\ln\gamma=-z^2\,A\,\sqrt[2]{I}$, with $A \!= \! 1.172$ M\textsuperscript{-\textonehalf}, which fits the first few data points up to a few mM, but not beyond. The factor \textit{I} is the ionic strength, in M, for a 1:1~salt the same as $c_{\infty,\text{M}}$, but in general $I=\text{\textonehalf} \sum_i {z_i}^2 c_i$. In the extended DH equation we divide by a term $1+a\s{DH}B\sqrt{I}$ where $a\s{DH}$ is an effective ion {diameter} and $B \!= \! 3.281$~M\textsuperscript{-\textonehalf}.nm\textsuperscript{-1}. This equation fits data for a radius $a\s{DH} \! = \! 0.46$~nm. We return to the description of the activity coefficient of NaCl and KCl in a larger concentration window in Appendix II. Another set of historical data, from 1932 and 1951, with HCl as the 1:1~salt, is described in Appendix III.

\begin{figure}
\centering
\hspace{5mm}
\includegraphics[width=1\textwidth]{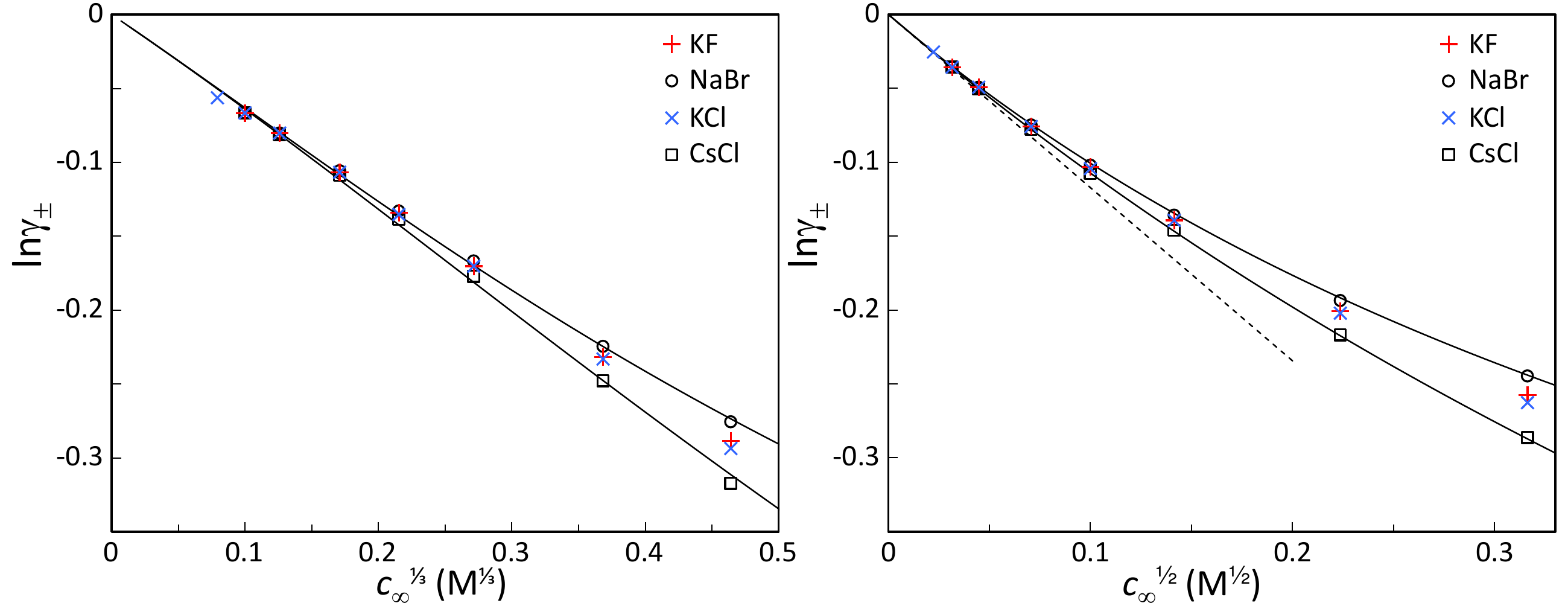}
\vspace{-14pt}
\caption{Data for $\ln\gamma_\pm$ of four 1:1 salts as function of $\sqrt[3]{c_\infty}$ in panel A, and as function of $\sqrt{c_\infty}$ in panel B. The lines in panel A are based on numerical evaluation of Eq.~\eqref{eq_ipe_full} for two values of the average ion radius ($\left\langle a \right\rangle \! = \! 0.18$ nm for the lower curve, $\left\langle a \right\rangle \! = \! 0.35$ nm for the upper curve). In panel B the dashed line is the DH limiting law and the solid lines are the extended DH theory that includes an effective ion diameter. }
\label{fig_lngamma_data}
\end{figure}

Next we discuss data of $\gamma_\pm$ of four 1:1 salt solutions from Hamer and Wu (1972), see Fig.~\ref{fig_lngamma_data}. When plotted against $\sqrt[3]{c_\infty}$, again these data do not extrapolate to the origin so we shift the data downward, all by the same amount of 0.031~{kT}. Then the data do follow the cube root law of Eq.~\eqref{eq_mu_11_practical} very well and we can improve the fit further when we numerically analyse Eq.~\eqref{eq_ipe_full} with the average ion radius as an adjustable parameter, for which we use $\left\langle a \right\rangle \! = \! 0.18$ nm for the lowest curve, and $\left\langle a \right\rangle \! = \! 0.35$~nm for the upper curve, i.e., for monovalent salts, the theoretical lines stay close to the cube root law irrespective of the value of the ion size, but they do have a clear dependence on ion size, corresponding to the variation in the data. The values of the ion sizes we must implement in the theory are very realistic. Interestingly, using the full calculation (see Appendix I), we have a decrease of $\ln \gamma$ with $c_\infty$ that is steeper than the analytical equation for low ion size, but more shallow at higher ion sizes. 
In Fig.~\ref{fig_lngamma_data}B we present the same data (without a vertical shift) as function of the square root of concentration. Beyond the first few points, the DH limiting law deviates from the data, but the extended DH equation fits data well with values of $a\s{DH}$ between 0.28~nm and 0.50~nm. 

It is quite unexpected that the data plotted in these graphs vs. the square root of concentration, all extrapolate well to the origin, but to do the same in graphs that plot data vs. a cube root dependence, they must be shifted downward by a small amount. 
%
The reason is that in the process to derive values for $\ln\gamma$ from experiments, a calibration, a theoretical extrapolation to $c_\infty \! \rightarrow \! 0$, is involved, because the measurement of chemical potential does not directly provide the reference values at zero salt concentration. Because the proposed laws (with $\ln\gamma_\pm$ dependent on $c_\infty$ in the dilute limit by a square root or cube root) lead to the slope of $\ln\gamma_\pm$ versus $c_\infty$ to diverge for $c_\infty \to 0$, we cannot extrapolate to zero salt concentration when the data are plotted versus the `linear' salt concentration (salt concentration raised to the power 1), see last row in Fig.~\ref{fig_benedek_villars_II}. Instead, we must extrapolate from the data (in a graph) with salt concentration raised to a power that is less than unity. And the choice of that power, either for instance 1/2 or 1/3, depends on what model one assumes is valid. So the calibration depends on the choice of what is assumed to be the correct limiting law. If this choice is wrong, reference values at zero salt concentration are not exactly determined, so there is quite some relevance to making sure the correct scaling is used. Because if the wrong scaling is used, the problem can only be remedied (though never very satisfactorily), by having more and more data at increasingly low salt concentration. Then the error is minimized. But it would have been better to have the right scaling in the dilute limit, because then we can robustly extrapolate (because there is then a linear dependence on salt concentration raised to that correct power), and we do not 
need to acquire and analyse data at ultralow salt concentrations. 

Thus, any theoretical extrapolation assumes a certain dependence of $\ln\gamma_\pm$ on concentration to make data extrapolate to the origin. Therefore, when a different underlying extrapolation theory is used, a group of data for $\ln\gamma_\pm$ will end up at somewhat 
different values (all shifted up or down by the same amount). For more details on this inherent problem in calibration, and the measurement of activity and osmotic coefficients in general, see Poirier (1966), Hamer and Wu (1972), Kuznetsova (2002), Marcus and Hefter (2006), and Malatesta \textit{et al.} (1994, 1997, 1999, 2020). In summary, the exact vertical position of a group of $\ln\gamma_\pm$-data is not obtained experimentally, 
%
and when published data correctly extrapolate to the origin when plotted against the square root of concentration, is not evidence of the correctness of this scaling law, but the assumption that this scaling law is valid, was part of the calibration procedure. 
As can be observed in Figs.~\ref{fig_lngamma_NaCl} and~\ref{fig_lngamma_data}, when we compare data for $\ln\gamma_\pm$ analysed based on a cube root calibration (panel A in both figures), versus a square root calibration (panel B in both figures), the \textit{data} come out somewhat differently (0.031~{kT} difference in Fig.~\ref{fig_lngamma_NaCl}). So the correct choice of scaling is very important unless we have extremely precise data at extremely low salt concentration.

\begin{figure}
\centering
\includegraphics[width=0.45\textwidth]{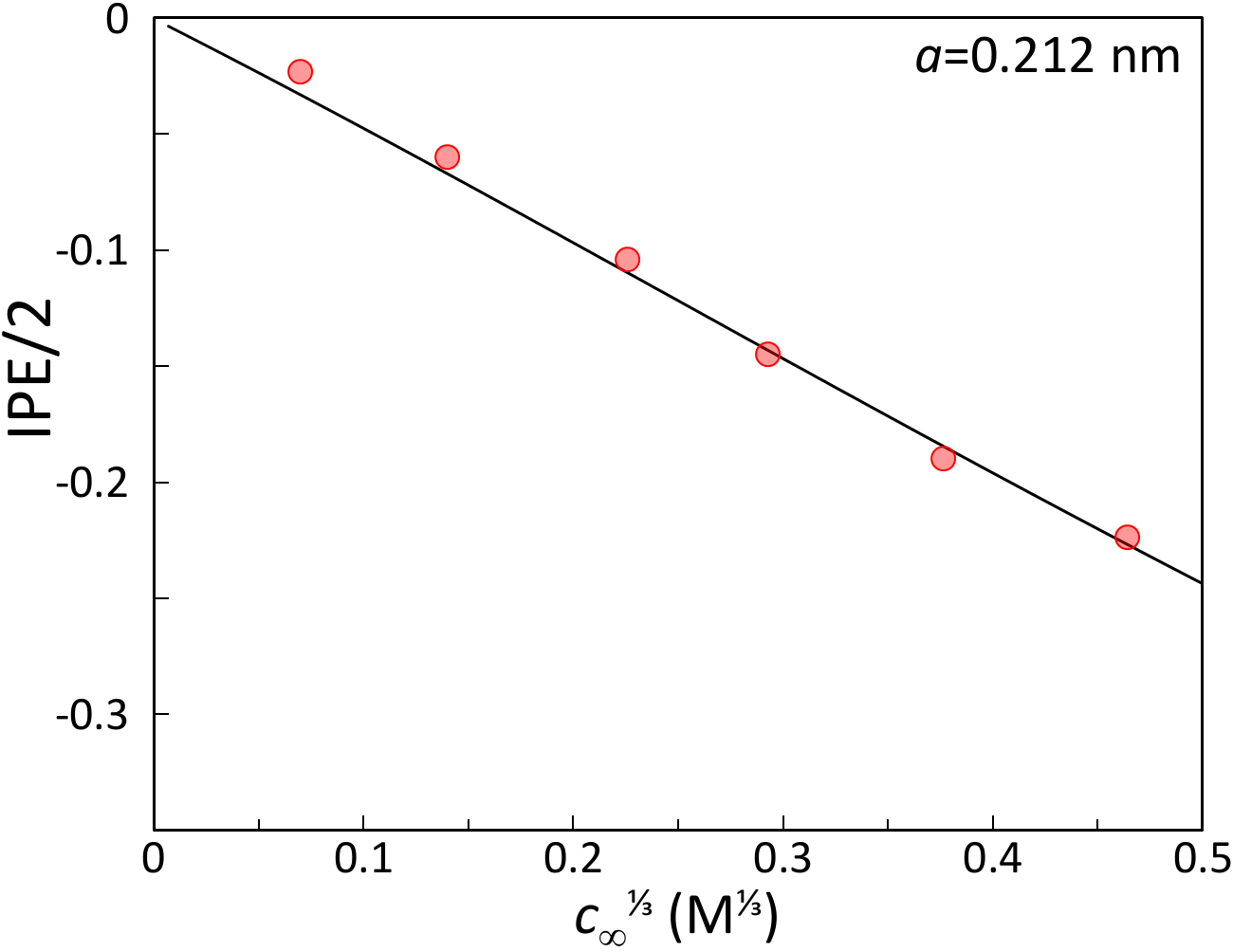}
\vspace{-0pt}
\caption{Simulation data by the restricted primitive model (RPM) from Lamperski (2007) of the electrostatic ion energy, that we compare with predictions of the 
ion pair model (solid line) for \textonehalf$\times$ the ion pair energy IPE for a salt with ion radius $a\!=\!0.212$~nm.}
\label{fig_IPE}
\end{figure}

\begin{figure}
\centering
\includegraphics[width=1\textwidth]{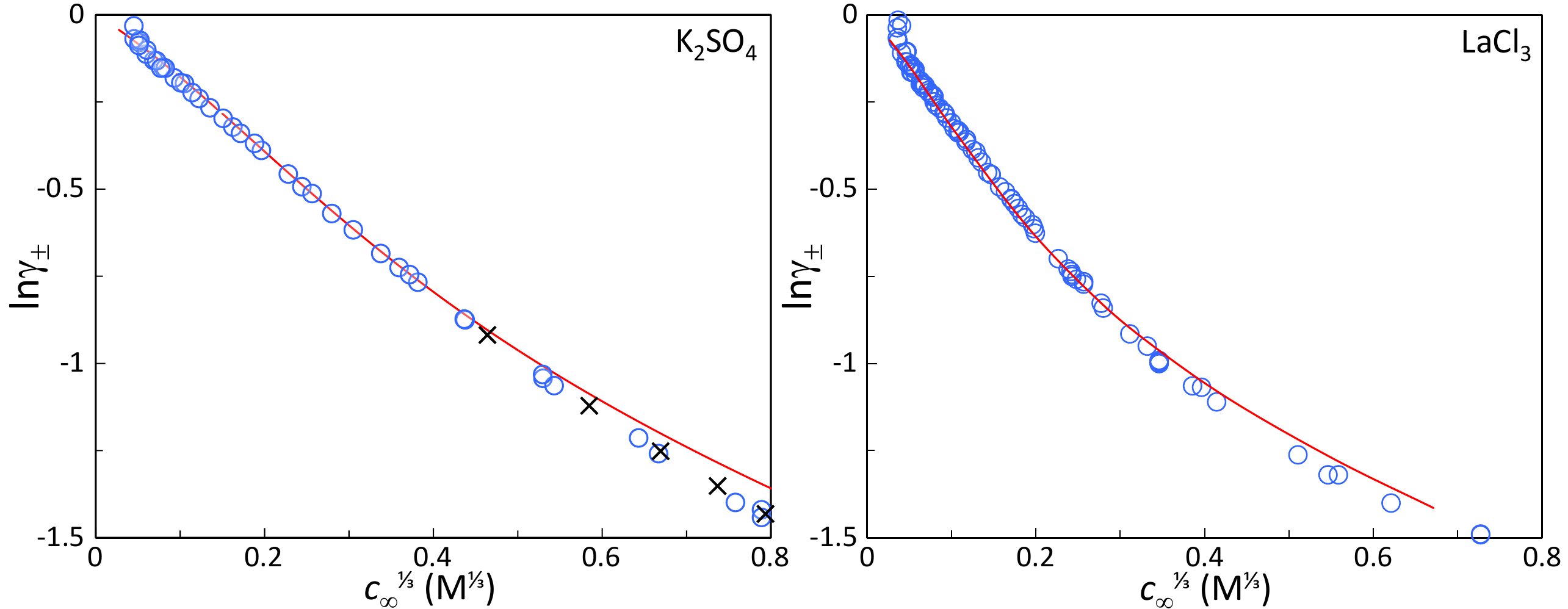}
\vspace{-8pt}
\caption{Data and theory for $\ln\gamma_\pm$ for a 2:1 and 3:1 salt, based on numerical analysis of the Coulombic interactions in an ionic triplet and quadruplet. For \ce{K2SO4}, the average radius is $\left\langle a \right\rangle \! = \! 0.18$ nm; for \ce{LaCl3}, $\left\langle a \right\rangle \! = \! 0.32$ nm. Up to 0.5 M\textsuperscript{1/3} for \ce{K2SO4} and up to 0.3 M\textsuperscript{1/3} for \ce{LaCl3} data follow a cube root dependence.}
\label{fig_asymmetric}
\end{figure}

\begin{figure}
\centering
\includegraphics[width=1\textwidth]{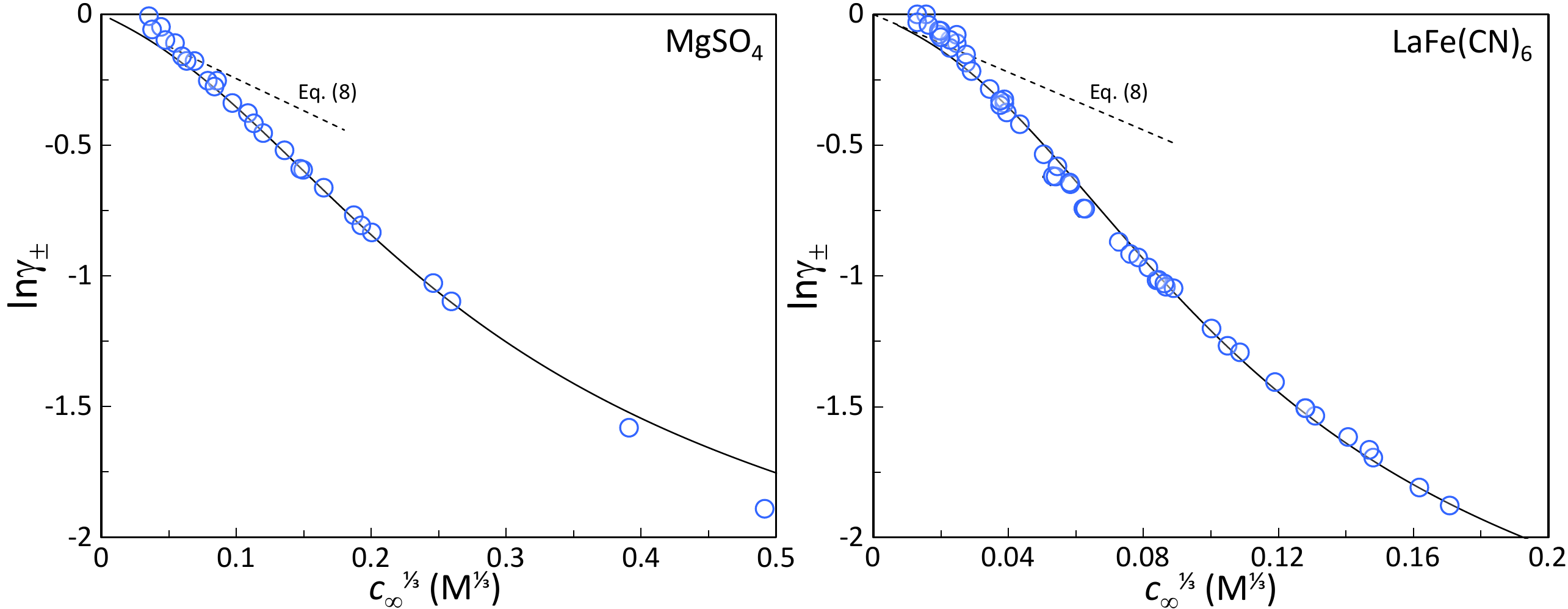}
\vspace{-9pt}
\caption{Data of $\ln\gamma_\pm$ of a divalent salt, \ce{MgSO4}, and a trivalent salt, \ce{LaFe(CN)6}, compared with numerical analysis, shown as lines, based on Eq.~\eqref{eq_ipe_full}. For \ce{MgSO4}, $\left\langle a \right\rangle \! = \! 0.25$ nm and for \ce{LaFe(CN)6}, $\left\langle a \right\rangle \! = \! 0.47$ nm. The upper dashed lines are the limiting law of Eq.~\eqref{eq_mu_11_practical}.}
\label{fig_lngamma_MgSO4_LaFeCN6}
\end{figure}

\begin{figure}
\centering
\includegraphics[width=0.55\textwidth]{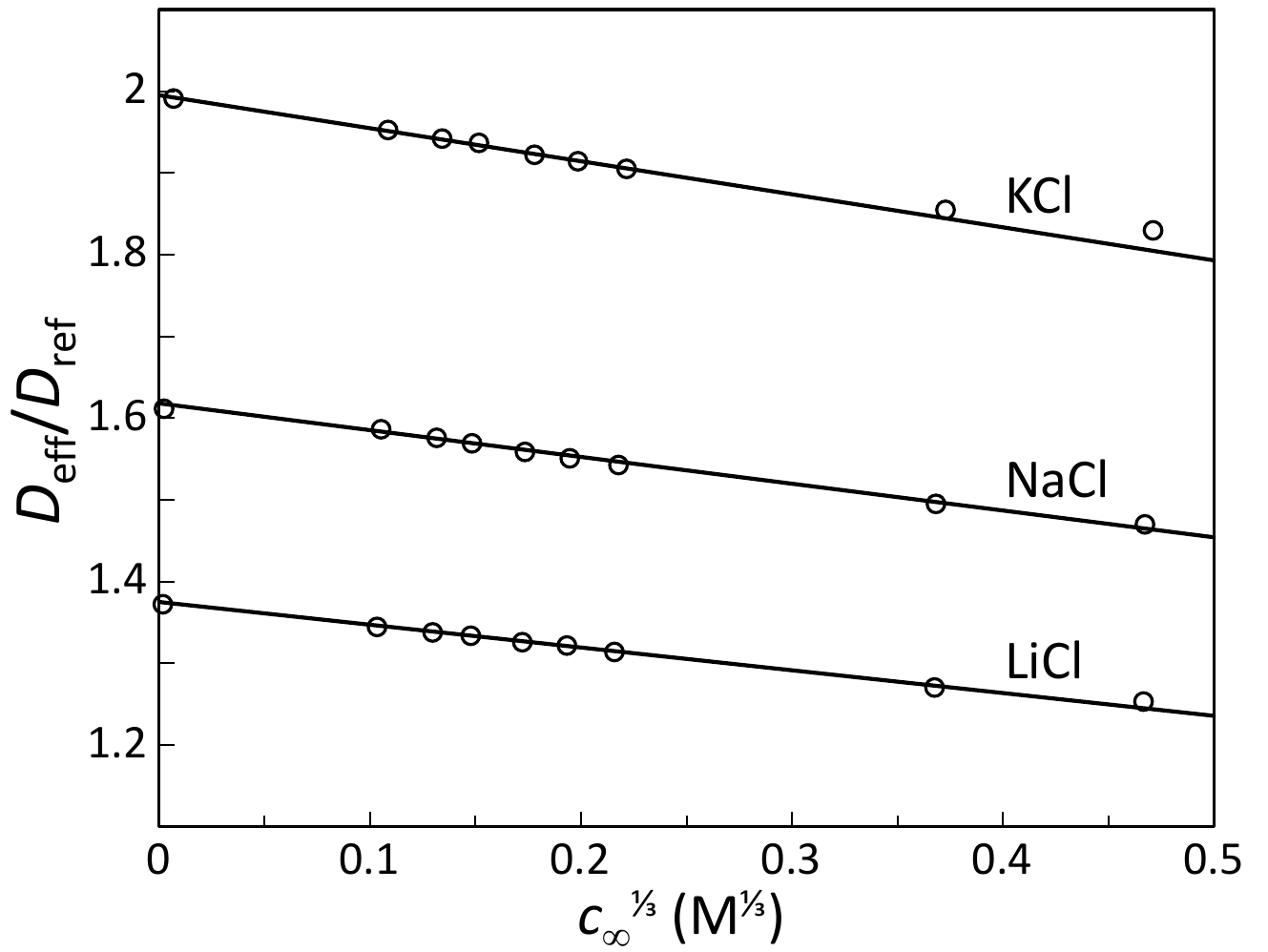}
\vspace{-7pt}
\caption{The effective diffusion coefficient for three salt solutions, based on Eq.~\eqref{eq_diffusion_nonideal}. The three lines are not parallel, but their slopes are proportional to $D\s{hm} = \left. D\s{eff}\right|_{c_\infty = 0}$ ($D\s{ref} \! = \! 1 \cdot 10^{-9}$~m\textsuperscript{2}/s).}
\label{fig_effective_diffusion}
\end{figure}

\begin{figure}
\centering
\includegraphics[width=0.55\textwidth]{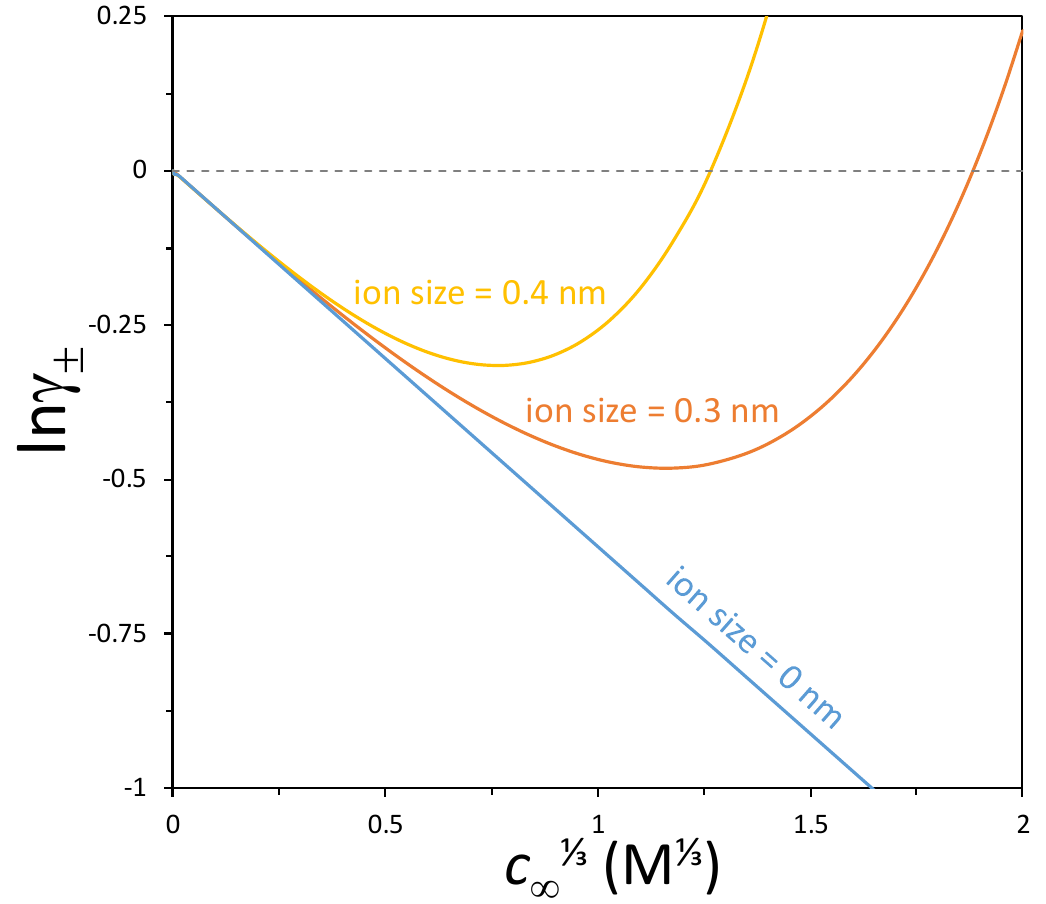}
\vspace{-8pt}
\caption{The activity correction of an ion, $\ln\gamma_{\pm}$, in a 1:1 salt solution, as function of the cube root of salt concentration. The two contributions considered are direct ion-ion Coulombic interactions, and the effect of volumetric exclusion. 
}
\label{fig_lngamma_th}
\end{figure}


Our calculation results can also be compared with results of simulations using the restricted primitive model (RPM) formalism. We use RPM data from Lamperski (2007) for a 1:1 salt with ions with a radius $a \!= \! 0.212$~nm, with $\varepsilon\s{r}\!=\!78.65$, which are reproduced in Fig.~\ref{fig_IPE}, together with additional RPM data kindly provided by S. Lamperski. We interpret these RPM results as the energy per ion pair, divided by 2, to be the energy per ion. They are not a contribution to an ion's chemical potential, so they are not the electrostatic contribution to $\ln\gamma_\pm$. This is logical, because to obtain calculation results for $\ln\gamma_\pm$ one must do RPM simulations at different values of $c_\infty$ and differentiate IPE (times $c_\infty$) with $c_\infty$. This is not the general procedure and thus a single RPM simulation run provides an electrostatic energy per ion. Indeed, if we would assume that RPM results such as in Fig.~\ref{fig_lngamma_NaCl}C are to be equated with $\ln\gamma_\pm$, they come out `too shallow', which must relate to the factor 4/3 that is different between IPE (divided by 2) and $\mu\s{i.i.c.i.}$ in the Bjerrum equation. 
Thus, we compare in Fig.~\ref{fig_IPE} RPM simulation results for the ion pair energy IPE, divided by 2, because this is how we interpret the RPM simulation data provided, and compare with calculation results from our model for the same ion radius of $a \! = \! 0.212$~nm. We find a very good agreement. 

We now continue with asymmetric salts such as a 2:1 salt and extend the theory for ion pairs to a theory for ion triplets. Let us assume the divalent ion is a cation, with ion radius \textit{a}\textsubscript{c}, which is at the center of a spherical coordinate system, 
and we have two monovalent anions that orbit around this center ion, $\alpha$ and $\beta$, each with 
radius $a_-$ and positional vectors $x_\alpha$ and $x_\beta$, counted from the center. The electrostatic energy of such a triplet, ITE, is calculated from
\begin{equation}
\text{ITE} = \frac{\int_0^{\pi} \int_0^{\pi/2} \int \int \overline{x}^2_\alpha \overline{x}^2_\beta \exp\left(-\psi\right) \psi  \sin \vartheta \text{d}\overline{x}_\alpha \text{d}\overline{x}_\beta \text{d}\vartheta \text{d}\phi}{\int_0^{\pi} \int_0^{\pi/2} \int \int \overline{x}^2_\alpha \overline{x}^2_\beta \exp\left(-\psi\right) \hphantom{\psi}\sin \vartheta  \text{d}\overline{x}_\alpha \text{d}\overline{x}_\beta \text{d}\vartheta \text{d}\phi}
\label{eq_ite}
\end{equation}
where the potential of a certain configuration is $\psi$, to be discussed below. The azimuthal angle between the vectors $x_\alpha$ and $x_\beta$ is $\phi$ and the polar angle is $\vartheta$. The magnitude of these position vectors is $\overline{x}_j$, and the integration over $\overline{x}_j$ runs between \textit{x}\textsubscript{min} and \textit{x}\textsubscript{max}, which we discuss below. These are the integration limits of the undefined $\int$'s in Eq.~\eqref{eq_ite}. 

We now have two questions to answer, first what are the integration limits \textit{x}\textsubscript{min} and \textit{x}\textsubscript{max} in this calculation, and second what is $\psi$? The minimum distance is the sum of the two radii, of divalent cation and monovalent anion. Below, we refer to $\left\langle a \right\rangle$ which is 
the average of these two radii. To find \textit{x}\textsubscript{max}, we use the same dependence on $c_\infty^{-\nicefrac{1}{3}}$ as used for the 1:1 salt, but because in this case the total ions concentration is three times the salt concentration, and not two times, in the numerical calculations we now use $3/2 \cdot n_\infty = 1/ \left(\alpha \cdot x\s{max}\right)^3$. In Eq.~\eqref{eq_ite}, the potential $\psi$ is a summation over three separate interactions, two attractive (between the cation and anion $\alpha$, and between the cation and anion $\beta$) and one repulsive, between $\alpha$ and $\beta$, each described by a modification of Eq.~\eqref{eq_corr_2}, namely each is given by $\psi = z_i z_j \lambda\s{B} / \widetilde{x}$ with the two \textit{z}'s the valencies of the two ions involved in each of the three interactions, and $\widetilde{x}$ the distance between the ions. For the interaction with the center ion, we simply have $\widetilde{x}=\overline{x}_j$, and for the anion-anion pair (which is a repulsion), this distance is given by $\widetilde{x}^2 = \overline{x}^2_\alpha + \overline{x}^2_\beta -2 \overline{x}_\alpha \overline{x}_\beta \sin \vartheta \cos \phi$.  
In this calculation the two anions can be overlapping, but that they are close is anyway unlikely as they are repulsive. We numerically analyse Eq.~\eqref{eq_ite} to obtain the energy of the ion triplet, multiply each energy with $c_\infty$ to obtain the free energy density, do this for many salt concentrations $c_\infty$, take the derivative with $c_\infty$, and divide by 3, to obtain $\ln \gamma_\pm$ where $\gamma_\pm$ is the mean activity coefficient of an ion in the 2:1 salt. A best fit to the data for \ce{K2SO4} from Malatesta \textit{et al.} (1999) is obtained for an average radius of $\left\langle a \right\rangle \! = \! 0.18$~nm, see Fig.~\ref{fig_asymmetric}A (data shifted down by 0.065~{kT}, blue circles). Also shown are data (black crosses) reported in the CRC Handbook of Chemistry and Physics, all shifted down by 0.1~{kT}. 


We can further extend the ion triplet model, now to a quadruplet model, to describe a 3:1 salt. The fourfold integral that was required for the 2:1 salt becomes a sevenfold integral to calculate the ion quadruplet energy, IQE, and for $x\s{max}$ we now use $4/2 \cdot n_\infty = 1/ \left(\alpha \cdot x\s{max}\right)^3$, and $\ln\gamma_\pm$ is now one-fourth of $\partial \left(n_\infty \cdot \text{IQE} \right) / \partial n_\infty$. Note that $\psi$ is now a summation over six separate ion-ion interactions, of which three involve the center ion and are attractive, and the other three interactions are repulsive, between the orbiting anions. All possible combinations of positions of the three anions that orbit the central cation are considered. We compare with data for \ce{LaCl3} obtained by Malatesta \textit{et al}. (1994) and we again obtain a very good fit; now we use $\left\langle a \right\rangle \! = \! 0.32$~nm (no offset for the data needed), see Fig.~\ref{fig_asymmetric}B. Thus the ion pair model can be successfully extended to a triplet model for asymmetric 2:1 salts, and to a quadruplet model for a 3:1 salt. Interestingly, the data follow a cube root scaling law, up to a concentration of 100 mM for the 2:1 salt and 30 mM for the 3:1 salt. The numerical model fits the data in this range and thus can be said to be in agreement with this trend. However, upon closer inspection, the output of the numerical model is that the slope of $\partial \ln \gamma_\pm / \partial \sqrt[3]{c_\infty}$ first becomes more negative and then increases again, i.e., the lines in Fig.~\ref{fig_asymmetric} first curve down and then up again. For practical purposes we can use Eq.~\eqref{eq_mu_11_practical} for these salts, and then for the 2:1 salt we have $b \! =\! 0.20$~mM\textsuperscript{-1/3}, and for the 3:1 salt, $b \! = \! 0.30$~mM\textsuperscript{-1/3}. Note that this is the mean activity coefficient, which we cannot separate out between anions and cations. 

We can also analyse theory and data for symmetric 2:2 and 3:3 salts. Malatesta and Zamboni (1997) present data for $\ln\gamma_\pm$ of \ce{MgSO4} which we reproduce in Fig.~\ref{fig_lngamma_MgSO4_LaFeCN6}A as function of $\sqrt[3]{c_\infty}$ (no offset). We can describe these data with Eq.~\eqref{eq_ipe_full} when we use an average ion radius of $\left\langle a \right\rangle \! = \! 0.25$~nm, which is the solid line in Fig.~\ref{fig_lngamma_MgSO4_LaFeCN6}A. The cube root limiting law (upper dashed line) of Eq.~\eqref{eq_mu_11_practical} is the mathematical limit of the full numerical theory based on Eq.~\eqref{eq_ipe_full} with $z\!=\!2$, but only applies at very low concentrations, up to $\sim \! 0.2$~mM. Thus, Eq.~\eqref{eq_mu_11_practical} does not describe the data. 
For a 2:2 salt, data for the osmotic coefficient are presented in Fig.~5 in Malatesta and Zamboni (1997) for \ce{ZnSO4}. At a concentration of 10~mM ($m\!= \! 10^{-2}$), the experimental value for $\phi$ is around 0.74, which is similar to results in Figs.~4 and 5 of Bjerrum (1926) for \ce{MgSO4}. We numerically calculate $\phi$ based on our theory and we come to exactly the same value. At 1~mM, Malatesta and Zamboni estimate $\phi \! \sim \! 0.9$, and this is also the value we calculate. 
For a 3:3 salt, data for the salt \ce{LaFe(CN)6} are presented in Fig.~\ref{fig_lngamma_MgSO4_LaFeCN6}B (Malatesta \textit{et al.}, 1995) with an offset of +0.09~{kT}, and with an average radius of $\left\langle a \right\rangle \! = \! 0.47$~nm they can be very well described by the numerical theory, Eq.~\eqref{eq_ipe_full}. The limiting law of Eq.~\eqref{eq_mu_11_practical}, evaluated for $z\!=\!3$, only agrees with the numerical theory up to a salt concentration of $\sim \! 10$~$\mu$M, so it can not be used to describe the data.  

How well would DH theory describe these data? For a symmetric $z \! : \! z$ salt, DH theory predicts $\ln\gamma$ to depend on \textit{z}\textsuperscript{2}, and in addition the ionic strength \textit{I} is a multiple \textit{z}\textsuperscript{2} of the salt concentration. If we plot the data for the 2:2 salt \ce{MgSO4} of Fig.~\ref{fig_lngamma_MgSO4_LaFeCN6}A against $\sqrt{c_\infty}$ and compare with the DH equations (offset in data 0.05~{kT}), they follow the DH square root law until around 10 mM, and beyond that point we get a reasonable fit of the extended DH equation with an effective diameter of $a\s{DH} \!= \! 0.3$~nm. A better fit would be when the initial slope would be increased by about 50\%, i.e., the numerator of the DH equation would need an additional factor of 1.5, and then the effective diameter would be $a\s{DH} \! = \! 0.60$~nm. For a 3:3 salt, the DH equation does not work at all, we need the numerator to be multiplied by a factor 1.7 to get the initial slope right at $a\s{DH} \! = \! 0$~nm, but when fitting the full curve this additional factor is ideally 2.1 and then $a\s{DH} \! = \! 2.0$~nm. Thus, in conclusion, the DH theory does not describe the activity coefficients in 2:2 salts very well, and not at all for 3:3 salt solutions.


In summary, the presented model based on the concept of ion ensembles works well for all salt solutions considered, symmetric and asymmetric. 
To describe ion-ion Coulombic interactions in an electrolyte solution, we only need to consider the interactions of ions with the few other ions that compensate its charge, which for a symmetric salt is one single ion of opposite charge. Why do we not have to consider the interactions with all other ions? Perhaps this is because the trajectories of all these other ions 
do not correlate with the positions of anions and cations inside the ion ensemble, i.e., anions and cations that are outside a given ion ensemble orbit around it without a preference to be near the anions or cations in the ensemble. And there are equal numbers of positive and negative charges in this outside solution, 
so all of their influences cancel. 


\underline{Effect on diffusion rates}. Next we briefly discuss the effect of the non-ideality $\ln\gamma$ on rates of diffusion. We solely discuss a 1:1 salt from this point onward. For transport of salt 
in the absence of electromigration or convection, we can combine regular diffusion with an activity correction, and obtain for the molar flux of a 1:1 salt 
\begin{equation}
J=-D\s{hm} \cdot \left(1 +  \frac{\partial \ln \gamma_\pm}{\partial \ln c_\infty}  \right) \cdot \frac{\partial c_\infty}{\partial x}
\label{eq_diffusion_nonideal}
\end{equation}
where $D\s{hm}$ is the harmonic mean diffusion coefficient of the salt (Biesheuvel and Dykstra, 2020). If we only consider the Coulombic effect discussed until now as a contribution to $\ln \gamma_\pm$, then Eq.~\eqref{eq_diffusion_nonideal} becomes
\begin{equation}
J=-D\s{eff} \frac{\partial c_\infty}{\partial x} \hspace{5mm},\hspace{5mm} D\s{eff}=D\s{hm} \cdot \left(1- \tfrac{1}{3} \cdot b \cdot \sqrt[3]{c_\infty} \right)
\label{eq_diffusion_nonideal_II}
\end{equation}
which 
can be compared with data for the effective diffusion coefficient $D\s{eff}$ versus concentration of various 1:1 salts given in Fig.~4 in Frank and Thompson (1959). We reproduce these data in Fig.~\ref{fig_effective_diffusion} and we observe that $D\s{eff}$ is linearly related to the product of $D\s{hm}$ and the cube root of salt concentration, exactly in line with Eq.~\eqref{eq_diffusion_nonideal_II}. Astonishingly, we exactly fit these three data sets with Eq.~\eqref{eq_diffusion_nonideal_II}, with only $D\s{hm}$ as fitting parameter, and $b$~as described above.\footnote{In this analysis, we have interpreted the `differential diffusion coefficient' in Frank and Thompson as $D\s{eff}$, i.e., the entire prefactor in front of $\partial c_\infty / \partial x$ in Eq.~\eqref{eq_diffusion_nonideal}.} 

Data for conductance in Ohm\textsuperscript{-1} in Fig.~3 of Frank and Thompson also show a linear dependence on $\sqrt[3]{c_\infty}$, but in this case the slope is the same for all salts, i.e., it is not proportional to the conductance in the dilute limit, and the slope is about 2$\times$ higher than for diffusion. This suggests that another analysis is required to describe the effect of Coulombic forces on conductivity. Also an effect of hindrance is expected, that reduces the diffusion coefficient at high concentrations, similar to that in sedimentation and fluidization of particles in water~(Biesheuvel \textit{et al.}, 2001). 


\underline{Ion volume effects}. Up until now we only considered Coulombic forces between ions as a contribution to $\ln \gamma$, as described for instance by Eq.~\eqref{eq_mu_11_practical}. 
In this last part we include a second contribution to $\ln \gamma$, which is due to ion volume exclusion, an effect which develops at moderate and high salt concentrations. To describe volume exclusion, we use the Carnahan-Starling equation of state, and take for anion and cation the same size, which results in
\begin{equation}
\mu_{\text{exc}}=\frac{3-\phi}{\left(1-\phi\right)^3}-3 = 8 \phi + 15 \phi^2 + \dots
\label{eq_CS}
\end{equation}
where volume fraction $\phi$ is ion volume times salt concentration times two.

We present in Fig.~\ref{fig_lngamma_th} curves of $\ln\gamma=\mu\s{i.i.c.i}+\mu_{\text{exc}}$ for a 1:1 salt as function of salt concentration and ion size, 
where for $\mu\s{i.i.c.i.}$ we use the Bjerrum equation, Eq.~\eqref{eq_mu_11_practical}. Ion size is thus only required in the calculation of the effect of volume exclusion. Fig.~\ref{fig_lngamma_th} 
shows how with increasing (hydrated) size of the ions, the curves bend upward more quickly with increasing salt concentration, and sooner reach values of $\ln\gamma$ above zero. 
When ions have different sizes, or even different shapes, extensions of the Carnahan-Starling equation are possible, 
but that kind of precision is probably not required. Instead, to describe thermodynamic properties of more concentrated solutions, it is likely 
more important to include interactions between ions such as (de-)protonation as function of pH, and salt 
pair formation. 

\medskip

In conclusion, we developed a theory that considers the Coulombic interactions between an ion and its countercharge. The range of separations of the ions is limited by a distance beyond which the neutral `outside' solution starts, and that distance depends on salt concentration. This theory can reproduce experimental data for 1:1 salts, as well as for 2:2, 3:3, 2:1 and 3:1 salts, and also describes data for the osmotic coefficient and diffusional rates. 

\section*{Appendix I: Detailed expressions for $\ln\gamma_\pm$ including ion size effects}

In Appendix I we extend Eq.~\eqref{eq_mu_11_practical} by evaluating more terms in Eq.~\eqref{eq_corr_4a} and making the required differentiation to calculate $\ln\gamma_\pm$ for a $z:z$ symmetric salt. Both the numerator, N, and denominator, D, in Eq.~\eqref{eq_corr_4a} are functions, $f\s{N}$ and $f\s{D}$, that depend on the minimum separation, $x\s{min}$, which is equal to twice the average ion radius $\left\langle a \right\rangle$, thus $x\s{min}=2 \left\langle a \right\rangle $, and depend on the salt concentration, $n_\infty$, according to $x\s{max}=1/\sqrt[3]{n_\infty}$, and $n_\infty = c_\infty \, N\s{av}$, with $\beta=z^2 \,\lambda\s{B}$. Thus, we can write the ion pair energy as
\begin{equation}
\text{IPE} = - \frac{f\s{N}}{f\s{D}} 
\end{equation}
where the two terms $f\s{N}$ and $f\s{D}$ can be read off from Eq.~\eqref{eq_ipe_full}. The contribution to the chemical potential because of Coulombic interactions is given by
\begin{equation}
\ln\gamma_\pm = \mu\s{i.i.c.i.}=\frac{1}{2} \,\frac{\partial}{\partial n_\infty}\left(n_\infty  \text{IPE}   \right) =- \frac{1}{2} \,\frac{\partial}{\partial n_\infty}\left(n_\infty \frac{f\s{N}}{f\s{D}}   \right) =- \frac{1}{2} \left( \frac{f\s{N}}{f\s{D}} + \frac{n_\infty}{f\s{D}} \cdot \left(\frac{\partial f\s{N}}{\partial n_\infty} - \frac{f\s{N}}{f\s{D}}\frac{\partial f\s{D}}{\partial n_\infty} \right)\right) \, .
\end{equation}

\noindent Both for $f\s{N}$ and $f\s{D}$ we have
\begin{equation}
\frac{\partial f\s{j}}{\partial n_\infty}=\frac{\partial f\s{j}}{\partial x\s{max}} \frac{\partial x\s{max}}{\partial n_\infty}
\end{equation}
and 
\begin{equation}
\frac{\partial x\s{max}}{\partial n_\infty} = - \tfrac{1}{3} x\s{max}^4 \, .
\end{equation}

\noindent Thus we have
\begin{equation}
\frac{\partial f\s{N}}{\partial n_\infty}= - \tfrac{1}{3} \cdot \exp \left(\beta/x\s{max} \right) \cdot x\s{max}^5 \cdot \beta 
\end{equation}
and
\begin{equation}
\frac{\partial f\s{D}}{\partial n_\infty}=- \tfrac{1}{3} \cdot \exp \left( \beta/x\s{max} \right) \cdot x\s{max}^6  \, .
\end{equation}
\noindent Thus, the differential terms, $\partial f\s{j} / \partial n_\infty$, are an explicit function of $x\s{max}$, thus of $n_\infty$. We arrive at
\begin{equation}
\ln\gamma_\pm  = - \frac{1}{2} \left( \frac{f\s{N}}{f\s{D}} - \tfrac{1}{3} \cdot \frac{n_\infty}{f\s{D}} \cdot \left( \beta  - \frac{f\s{N}}{f\s{D}}  \cdot x\s{max}  \right) \cdot x\s{max}^5 \cdot \exp \left(\beta/x\s{max} \right) \right)\, .
\end{equation}

\noindent We write the integrated functions, $f\s{N}$ and $f\s{D}$, as a series of terms, with $f\s{N}$ given by
\begin{equation}
f\s{N}  = \left.\vphantom{\frac{1}{i! }} \tfrac{1}{2} \beta x^2 \right] + \left.\vphantom{\frac{1}{i! }} \beta^2 x \right] + \left.\vphantom{\frac{1}{i! }}  \tfrac{1}{2} \beta^3 \ln{x} \right] - \sum_{i=4 \dots \infty} \left. \frac{1}{\left(i-1\right)! \cdot \left(i-3\right)} \cdot \beta^i \cdot x^{3-i}\right]  
\end{equation}
where each operation `]' refers to evaluation of the argument at the upper and lower limits, $x\s{max}$ and $x\s{min}$, and taking the difference. The notation `!' refers to the factorial function, such as $3! = 1 \cdot 2 \cdot 3$. For $f\s{D}$ we have
\begin{equation}
f\s{D}  = \left.\vphantom{\frac{1}{i! }} \tfrac{1}{3} x^3 \right] + \left.\vphantom{\frac{1}{i! }} \tfrac{1}{2} \beta x^2 \right] + \left.\vphantom{\frac{1}{i! }} \tfrac{1}{2} \beta^2 x \right] + \left.\vphantom{\frac{1}{i! }} \tfrac{1}{6} \beta^3 \ln{x} \right] - \sum_{i=4 \dots \infty} \left. \frac{1}{i! \cdot \left(i-3\right)} \cdot  \beta^i \cdot x^{3-i}\right]  \, .
\end{equation}
These equations can be programmed easily in commercial spreadsheet software, and an example sheet is available from the author. All curves in all graphs in this paper for symmetric salts can be calculated explicitly as function of $c_\infty$ with the above equations. For a 3:3 salt, we must evaluate up to 15 terms for the calculation to converge (i.e., only then additional terms no longer change the outcome), and to have agreement with full numerical results. For 2:2 salts, we must evaluate at least 10 terms. But for a 1:1 salt, a much lower number suffices, with two terms already resulting in close agreement to the numerical result, and three terms agreeing very closely. 
%
%

\noindent If we only use the first term in each series of $f\s{N}$ and $f\s{D}$, we obtain
\begin{equation}
\ln\gamma_\pm = - \frac{1}{2} \cdot \left[ \frac{\left. \tfrac{1}{2} \beta x^2 \right] }{\left.\tfrac{1}{3} x^3 \right] } - \tfrac{1}{3} \cdot \frac{n_\infty}{\left.\tfrac{1}{3} x^3 \right] } \cdot \left( \beta  - \frac{\left.\tfrac{1}{2} \beta x^2 \right] }{\left.\tfrac{1}{3} x^3 \right] }  \cdot x\s{max}  \right) \cdot x\s{max}^5 \cdot \exp \left(\beta/x\s{max} \right) \right]
\end{equation}
%
%
which we can express in $n_\infty$ instead of $x\s{max}$, 
resulting in
\begin{equation}
\ln\gamma_\pm = - \frac{3}{4} \cdot \frac{\beta}{1-8\left\langle a \right\rangle^3 n_\infty} \cdot \left[  \sqrt[3]{n_\infty}-4\left\langle a \right\rangle^2 n_\infty -  \left(  \tfrac{2}{3} \cdot \sqrt[3]{n_\infty} -   \frac{\sqrt[3]{n_\infty}-4\left\langle a \right\rangle^2n_\infty}{1-8 \left\langle a \right\rangle^3 n_\infty}\right)   \cdot \exp \left(\beta \sqrt[3]{n_\infty} \right) \right]
\end{equation}
%
%
%
for which the series expansion for $n_\infty \to 0$ is
\begin{equation}
\ln\gamma_\pm = - \beta \cdot   n_\infty^{1/3} -  \tfrac{1}{4} \cdot \beta^2 \cdot n_\infty^{2/3} + \tfrac{3}{4} \cdot \beta \cdot \left(8 \cdot \left\langle a \right\rangle^2 -\tfrac{1}{6} \cdot \beta^2 \right) \cdot n_\infty + \mathcal{O}\left(n_\infty^{4/3} \right)
\label{eq_Bjerrum_extended_formal}
\end{equation}
where the first term is the same as Eq.~\eqref{eq_ipe}. The second term shows how with increasing salt concentration, $\ln\gamma_\pm$ decreases faster than linearly, exactly as the theory line in Fig.~\ref{fig_lngamma_data}A for $\left\langle a \right\rangle=0.18$~nm. The average ion radius $\left\langle a \right\rangle$ first shows up in the third term, which is linear in $n_\infty$. For a 1:1 salt the ion size effect is positive when $\left\langle a \right\rangle > \lambda\s{B} / \left(4 \sqrt{3}\right)$, and this occurs for $\lambda\s{B}\!=\!0.716$~nm when $\left\langle a \right\rangle > 0.103$~nm. In reality ions plus their hydration shell are always larger than that minimum value. So Eq.~\eqref{eq_Bjerrum_extended_formal} illustrates that for a 1:1 salt an ion size effect develops which 
moves the curves of $\ln\gamma_\pm$ versus $c_\infty$ upward with increasing ion size, in line with results presented in Fig.~\ref{fig_lngamma_data}A. 

We can rewrite this last equation to a dependence on salt concentration $c_\infty$ in mM which results in 
\begin{equation}
\ln \gamma_\pm =  - b \cdot c_\infty^{1/3} -\tfrac{1}{4} \cdot b^2 \cdot c_\infty^{2/3}  + 6 \cdot b^3 \cdot q  \cdot c_\infty
\label{eq_bjerrum_extended}
\end{equation}
where as in the main text $b =  \, z^2 \, \lambda\s{B} \, \sqrt[3]{N\s{av}}$, where $\lambda\s{B}$ is the Bjerrum length, which at room temperature is $b \! = \!  0.0605\text{~}\left(\text{mM}\right)^{-1/3}$. Eq.~\eqref{eq_bjerrum_extended} is what we refer to as the extended Bjerrum equation. 
To simplify notation, we introduce here a dimensionless size factor, $q$, that depends on the average ion radius, $\left\langle a \right\rangle$, according to $q=\left(\left\langle a \right\rangle / \lambda\s{B} \right)^2 - 1/48$. Eq.~\eqref{eq_bjerrum_extended} works very well for 1:1 salts as we will show in Appendix II for \ce{NaCl} and \ce{KCl}. 

When we use Eq.~\eqref{eq_bjerrum_extended} for a 2:2 salt, the valency \textit{z} is a factor 2 larger, and thus \textit{b} is 4~times larger. However, it then only in first approximation agrees with the data and full calculations. Instead, an accurate fit for a 2:2 salt is obtained when in addition to the four-times larger \textit{b}, the second term in Eq.~\eqref{eq_bjerrum_extended} is multiplied by 8, and for \textit{q} we use $q \! = \! 0.23$.




\section*{Appendix II: Extended dataset for NaCl and KCl}

\begin{figure}
\centering
\includegraphics[width=1\textwidth]{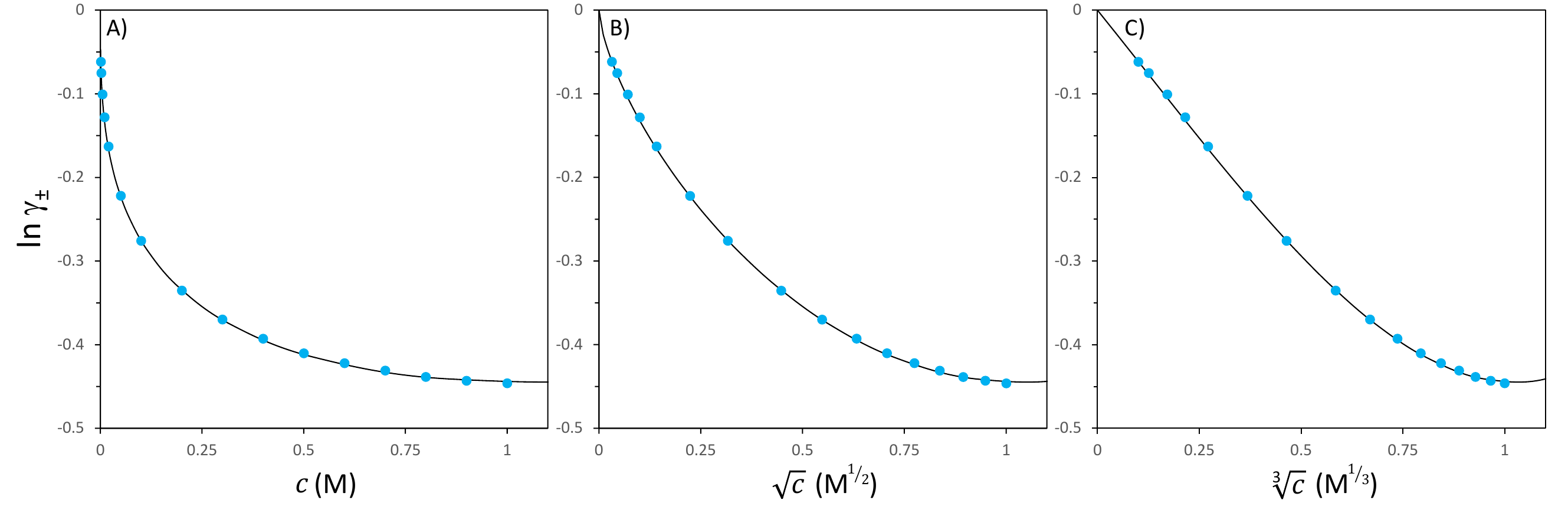}
\vspace{-20pt}
\caption{Data of activity coefficient of NaCl solution, $\ln\gamma_\pm$, up to a salt concentration of $c_\infty \! = \! 1$~M, and compared with the extended Bjerrum equation, Eq.~\eqref{eq_bjerrum_extended}, using a size factor $q \! = \! 0.19$.}
\label{fig_NaCl_extended}
\end{figure}

We evaluate the extended Bjerrum equation, Eq.~\eqref{eq_bjerrum_extended}, by comparing with data for \ce{NaCl} and \ce{KCl} up to a salt concentration of 1 M, thus extending to higher salt concentrations than in the earlier analysis reported in Figs.~\ref{fig_lngamma_NaCl} and~\ref{fig_lngamma_data}. As Fig.~\ref{fig_NaCl_extended} shows, the extended Bjerrum equation accurately describes the data for NaCl. For an optimal fit we use a dimensionless size factor of $q \! = \! 0.19$, which recalculates to an average ion radius of $\left\langle a \right\rangle \! = \!  0.33$~nm, similar to the average ion radius used for NaBr in Fig.~\ref{fig_lngamma_data}. As can be clearly observed in the plot of $\ln\gamma_\pm$ versus the square root of concentration (panel B), there simply is no evidence of a linear scaling of $\ln\gamma_\pm$ with the square root of concentration, $\sqrt{c_\infty}$, in the dilute limit. However, when the data are plotted against the cube root of concentration (panel C), the scaling of (i.e., the linear dependence of) $\ln\gamma_\pm$ on the cube root of concentration, $\sqrt[3]{c_\infty}$, is very clear over a large concentration range, and indicative of the fact that indeed the cube root scaling is the correct scaling in the dilute limit, even up to quite high concentrations of approx. 0.1~M, a scaling relationship that was already identified by Bjerrum in 1916 and 1919.

\begin{figure}
\centering
\includegraphics[width=0.55\textwidth]{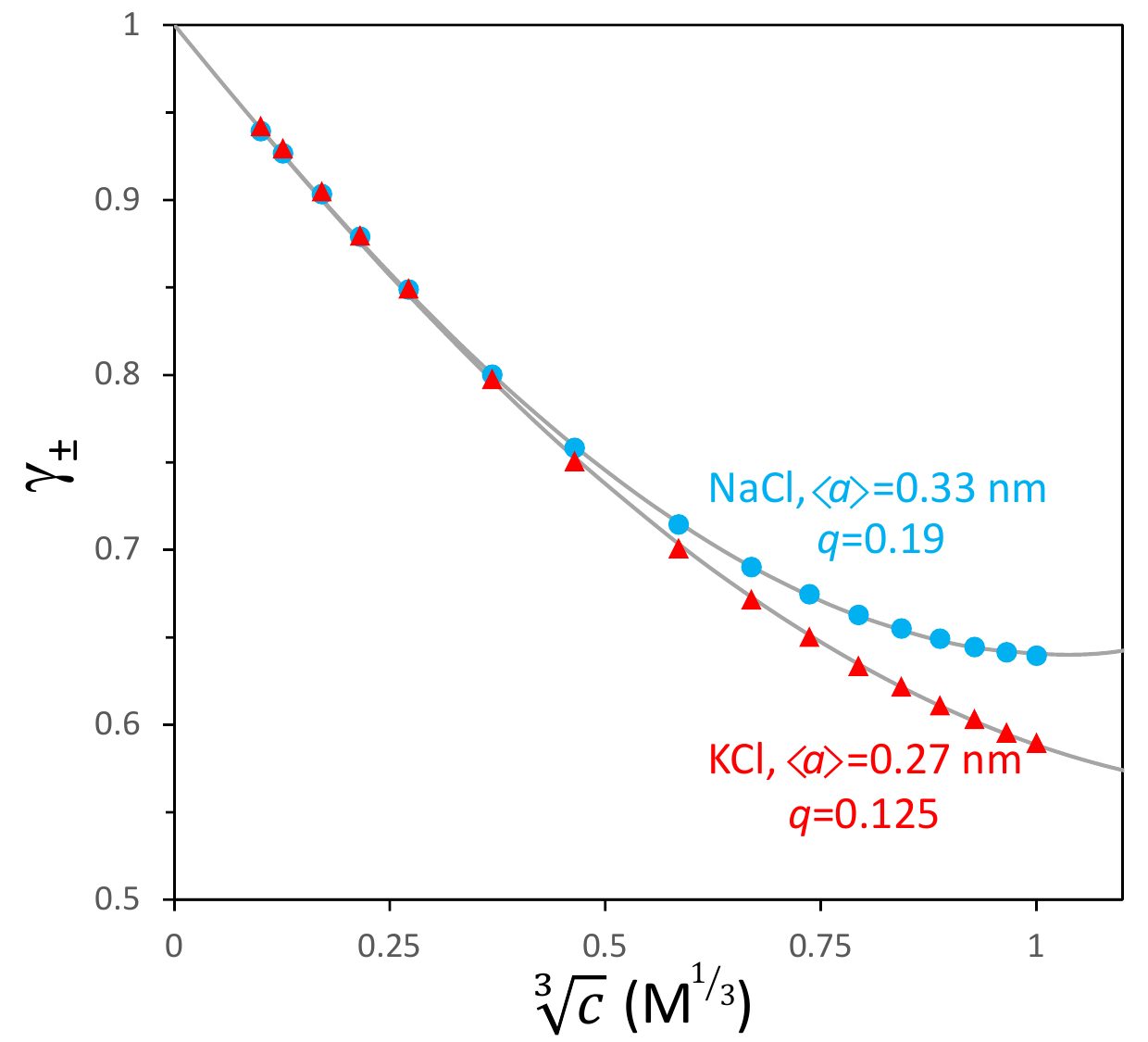}
\vspace{-8pt}
\caption{Mean activity coefficient, $\gamma_\pm$, for \ce{KCl} and \ce{NaCl} salt solutions fitted with the extended Bjerrum equation, Eq.~\eqref{eq_bjerrum_extended}, with only the size factor \textit{q} different for the two salts ($q\!=\!0.125$ for \ce{KCl} and $q\!=\!0.19$ for \ce{NaCl}).}
\label{fig_gamma_KCl_NaCl}
\end{figure}

In Fig.~\ref{fig_gamma_KCl_NaCl} we plot $\gamma_\pm$ for \ce{KCl} and \ce{NaCl} with only the size factor \textit{q} different (smaller for KCl). This plot is related to Fig.~3.26 in Bockris and Reddy (1970). 
Though the size factors \textit{q} are about 50\% different ($q\!=\!0.125$ for \ce{KCl} and $q\!=\! 0.19$ for \ce{NaCl}), the average ion radius $\left\langle a \right\rangle$ is not that different, for \ce{KCl} it is $\left\langle a \right\rangle \! = \! 0.27$~nm and for \ce{NaCl} it is $\left\langle a \right\rangle \! = \! 0.33$~nm. Fig.~\ref{fig_gamma_KCl_NaCl} shows that the two datasets overlap at a low salt concentration, and only beyond $\sim \! 0.1$~M do they start to diverge from one another, with the activity coefficient lower for the salt with the smaller average ion radius.

\section*{Appendix III: Data for HCl from Carmody (1932) and Hills and Ives (1951)}

\begin{figure}
\centering
\includegraphics[width=1\textwidth]{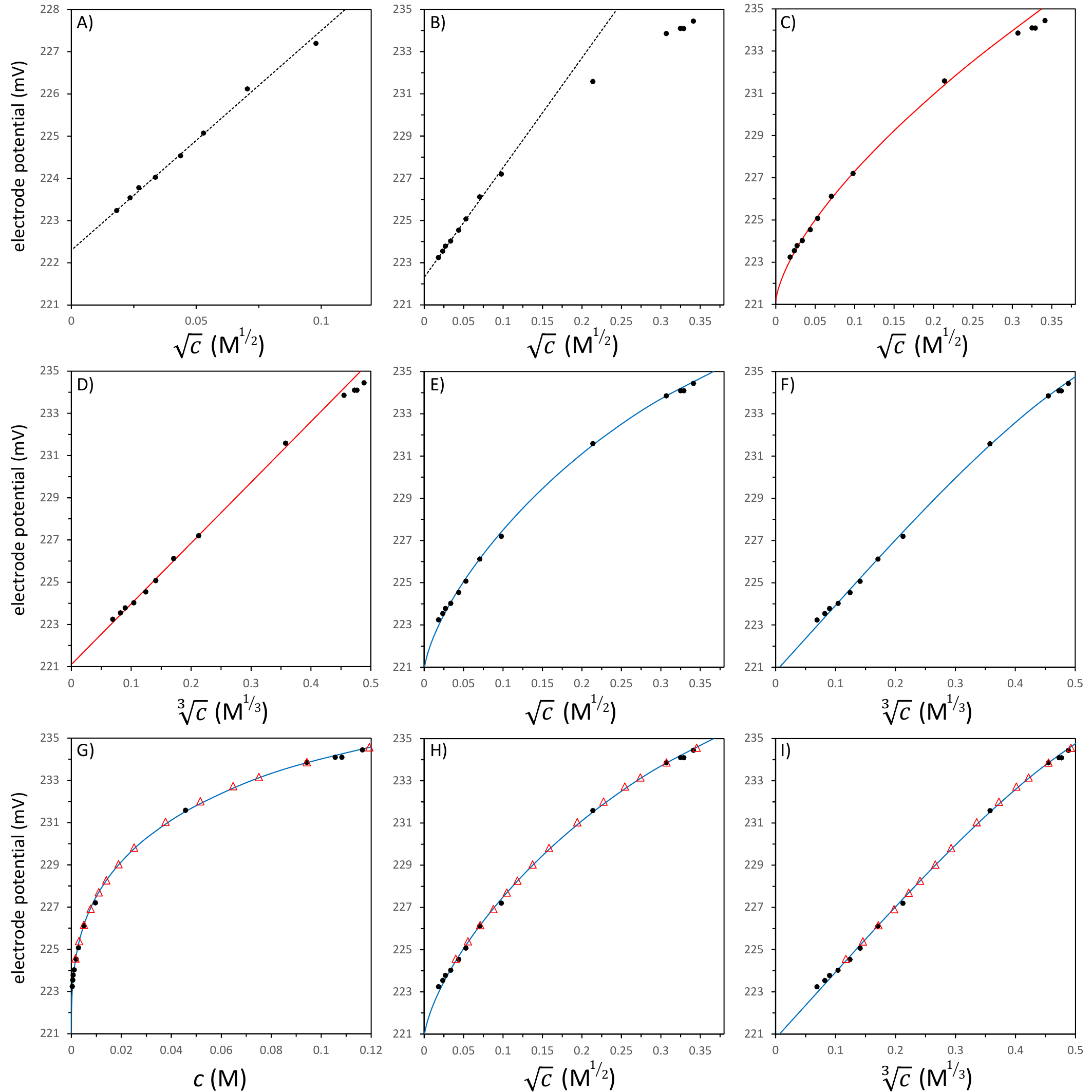}
\vspace{-8pt}
\caption{Electrode potential corrected for Nernst concentration effect, from Carmody (1932), as described in Dole (1935), as function of HCl concentration, \textit{c}. Electrode potential can be recalculated to $\ln\gamma_\pm$ after subtracting a reference value $E_0$, taking the negative, and dividing by $2\cdot RT/F$. A). Data as presented in Dole (1935) and later literature, plotted as function of square root of HCl concentration, $\sqrt{c}$. In these sources only 8 datapoints are reported. B). Same as A) but extended from 8 to all 13 datapoints that were originally reported by Carmody. The dashed line describes a scaling of $\ln\gamma_\pm$ with $\sqrt{c}$ C). Fit of data assuming a cube root dependence of $\ln\gamma_\pm$ on HCl concentration, in accordance with the Bjerrum theory, plotted against $\sqrt{c}$. D). Same as C) but plotted as function of the cube root of concentration, $\sqrt[3]{c}$. E). As in D) but now fitted with the extended Bjerrum theory, Eq.~\eqref{eq_bjerrum_extended}, plotted against $\sqrt{c}$, and F) plotted against $\sqrt[3]{c}$. G-I). Including data for $\ln\gamma_\pm$ from Hills and Ives (1951) (red triangles) and plotted versus \textit{c}, $\sqrt{c}$, and $\sqrt[3]{c}$.}
\label{fig_benedek_villars}
\end{figure}

A dataset that is often reported is for HCl solutions, with an electrode potential corrected for an ideal Nernst concentration effect, 
as function of HCl concentration. This dataset was obtained by Carmody (1932) and analysed in detail in the textbook of Dole (1935). 
We encountered this dataset in Benedek and Villars (2000). 
From the dataset we can derive $\ln \gamma_\pm$ by first subtracting an assumed reference value for the electrode potential at $c\!=\!0$, i.e.,~$E_0$, then taking the negative and dividing by $2 \cdot RT /F$. But first, in Fig.~\ref{fig_benedek_villars}A these data are reproduced as they are presented in Dole (1935), and in Benedek and Villars (2000).\footnote{However, it must be noted that in Benedek and Villars, one crucial datapoint that deviated the most from the square-root scaling was erroneously plotted to now perfectly match the square-root scaling. In Carmody this datapoint is (4.965~mM,~226.12~mV), and it is plotted by Benedek and Villars at approx. (5.80~mM,~226.05~mV).} 
Presented in this manner, the data seem to scale fairly well 
with the square root of HCl concentration, $\sqrt{c}$. However, five out of the thirteen original datapoints in Carmody are not included in this analysis of Dole, and later the same in the analysis by Benedek and Villars. 
%
%
If we also include these five data points that were left out, panel B is obtained but now the data no longer support a 
scaling of the potential with $\sqrt{c}$. In panel C the Bjerrum equation, Eq.~\eqref{eq_mu_11_practical}, is used with data and theory still plotted against $\sqrt{c}$, and in panel D against the cube root of concentration, $\sqrt[3]{c}$. The Bjerrum equation fits data much better than an assumed proportionality with $\sqrt{c}$. The slope in the curve, assuming $RT/F\!=\! 25.69\text{~mV}$, a value also used by Carmody and the later authors, results in a prefactor of $b = -0.56\text{~M}^{-1/3}$, which is about 5\% lower than $-0.58\text{~M}^{-1/3}$, a number mentioned in the main text based on analysis of experiments by Bjerrum. 
%
%
In panels E and F we compare data with the extended Bjerrum equation, Eq.~\eqref{eq_bjerrum_extended}, as function of $\sqrt{c}$ and $\sqrt[3]{c}$, respectively. We can use $b=-0.605\text{~M}^{-1/3}$ as generally used in the main text, in combination with a size factor of $q\!=\!0.32$. Now all data are fitted well, over the entire concentration range from 0.33 to 125 mM. That significant progress has been achieved, can be observed from comparing the previous analysis (panel A, published in the textbook of Benedek and Villars in 2000), and the new result, panel F. It is of great interest to notice that dependent on the chosen equation for $\ln \gamma_\pm$ 
different values are derived for $E_0$, from 222.3~mV for a $\sqrt{c}$ scaling (see Fig.~\eqref{fig_benedek_villars}A,B), to 221.1~mV for a $\sqrt[3]{c}$ scaling with the Bjerrum equation (panels C,D), and 220.8~mV for the extended Bjerrum equation (panels E-F). In panels G-I we also include data from Hills and Ives (1951), Table III, that we discuss next, recalculated to electrode potential by multiplying with $-2 \, RT/F$ (with $RT/F\!=\!25.69$~mV as used above), and adding $E_0\!=\!222.3$~mV. 

\begin{figure}
\centering
\includegraphics[width=1\textwidth]{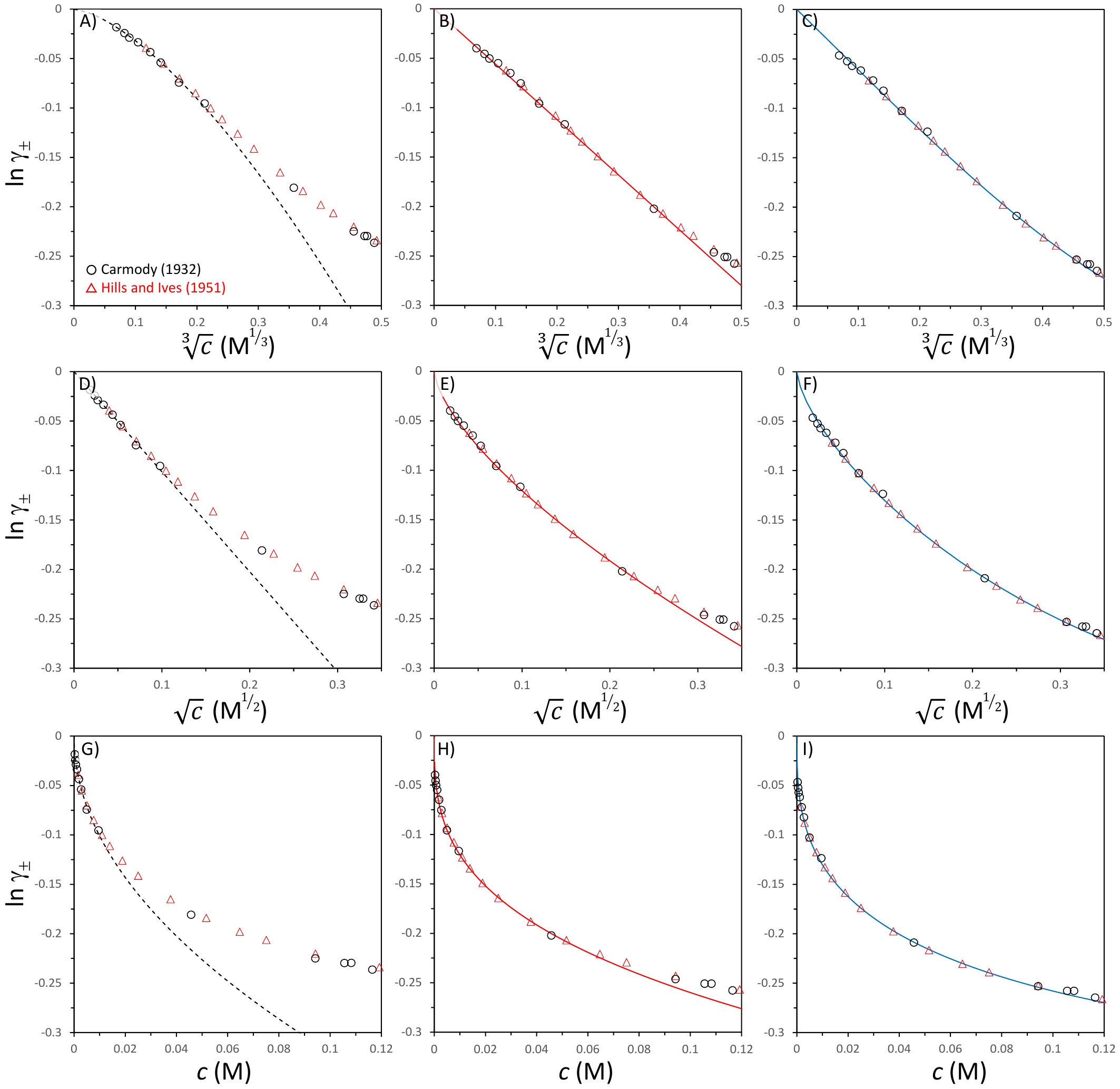}
\vspace{-8pt}
\caption{Mean activity coefficient of HCl solutions according to data by Carmody (1932) and Hills and Ives (1951) for HCl at $25\text{~}^{\circ}\text{C}$, as function of the concentration of acid, $c$. A-C) $\ln\gamma_\pm$ plotted against the cube root of concentration, $\sqrt[3]{c}$; D-F) $\ln\gamma_\pm$ plotted against $\sqrt{c}$; and G-I) $\ln\gamma_\pm$ plotted against \textit{c}. In the first column, the dashed black lines are based on a scaling of $\ln\gamma_\pm$ with $\sqrt{c}$, in the second column, red lines use the Bjerrum equation, thus are based on $\ln\gamma_\pm$ scaling with $\sqrt[3]{c}$, and in the third column, blue lines are based on the extended Bjerrum equation, Eq.~\eqref{eq_bjerrum_extended}.}
\label{fig_benedek_villars_II}
\end{figure}

Additional data of $\ln\gamma_\pm$ of HCl in water obtained by Hills and Ives (1951) are now also considered, again obtained from the measurement of electrode potentials. We present them in Fig.~\ref{fig_benedek_villars_II}, together with the data of Carmody (1932). The two datasets overlap perfectly after a slight shift in vertical position (of an entire dataset), which as discussed before, can always be done. 
In panels A-C, these data are plotted as function of $\sqrt[3]{c}$, as function of $\sqrt{c}$ in panels D-F, and versus concentration \textit{c} in M in panels G-I. In the first column the theory lines (dashed black lines) assume a scaling of $\ln\gamma_\pm$ with $\sqrt{c}$, in the second column, the red lines assume the Bjerrum scaling with $\sqrt[3]{c}$, and in the last column blue lines are based on the extended Bjerrum equation, Eq.~\eqref{eq_bjerrum_extended}. 
In each column data have a different `up/down shift' to align with the respective (scaling) law, but the maximum difference is small, $\sim 0.03$~{kT}. We notice that a scaling of $\ln\gamma_\pm$ with the square root of concentration (dashed lines in left column), does not fit data very well. Using the Bjerrum equation that is based on a scaling with the cube root of concentration, Eq.~\eqref{eq_mu_11_practical}, we obtain a perfect fit up to around a concentration of 80 mM, see the red lines in the middle column. In this analysis, we use for the prefactor \textit{b} the value earlier mentioned, of $b \!= \! 0.56\text{~M}^{-1/3}$. When we use the extended Bjerrum equation, Eq.~\eqref{eq_bjerrum_extended}, the entire dataset can be fitted almost perfectly, as shown by the blue lines in the rightmost column. And in this last case we can use the value of \textit{b} that was also used in the main text, of $b \!= \! 0.605\text{~M}^{-1/3}$, and for the size factor we have $q\!=\!0.32$. The extended Bjerrum equation, as used in the rightmost column (blue lines), results in the best fit to the data of the three theoretical models that were analysed.

\section*{Appendix IV: Osmotic pressure of NaCl and KCl solutions}

Based on the expression for $\ln\gamma_\pm$ given by Eq.~\eqref{eq_bjerrum_extended}, we can also derive an accurate expression for the osmotic pressure, $\Pi$, for which the Gibbs-Duhem equation can again be used, just like in the derivation of Eq.~\eqref{eq_osm_pressure_cell_model_11}. For a 1:1 salt we then arrive at an expression for the osmotic coefficient, $\phi$, given by
\begin{equation}
\phi = \frac{\Pi}{\Pi\s{id}} =  1 - \tfrac{\overline{b}}{4} \cdot \overline{c}_\infty^{1/3} -\tfrac{1}{10} \cdot \overline{b}^2 \cdot \overline{c}_\infty^{2/3}  + 3 \cdot \overline{b}^3 \cdot q  \cdot \overline{c}_\infty = 1 - \overline{\gamma} \cdot \overline{c}_\infty^{1/3} -\tfrac{8}{5} \cdot \overline{\gamma}^2 \cdot \overline{c}_\infty^{2/3}  + 192 \cdot \overline{\gamma}^3 \cdot q  \cdot \overline{c}_\infty 
\label{eq_osmotic_pressure_extended}
\end{equation}
where the ideal osmotic pressure is $\Pi\s{id}=2  c_\infty RT$. Concentrations here have an overline-notation to imply they are dimensionless, i.e., $\overline{c}=c/c\s{ref}$, with $c\s{ref}=1$~mM. In this way, constants $\overline{b}$ and $\overline{\gamma}$ are also dimensionless, with $\overline{b}=0.0605$ and $\overline{\gamma}=0.0151$. The two notations, one expressed in $\overline{b}$, the other in $\overline{\gamma}$, are equivalent. Results of Eq.~\eqref{eq_osmotic_pressure_extended} are presented in Fig.~\ref{fig_osm_coeff} for NaCl and KCl. The size factor \textit{q} was derived for NaCl and KCl in Appendix I based on comparison with data for $\gamma_\pm$, see Fig.~\ref{fig_gamma_KCl_NaCl}, to be $q \! = \! 0.19$ for NaCl and $q \! = \! 0.125$ for KCl. In Fig.~\ref{fig_osm_coeff}, we show the very good fit to data of the equation for the osmotic coefficient, Eq.~\eqref{eq_osmotic_pressure_extended}, up to a salt concentration of approx. 1.5 M.

It is interesting to evaluate the data for the osmotic coefficient, $\phi$, as provided by Hamer and Wu (1972)~\cite{Hamer_Wu}. The interesting point is that it turns out that these data almost exactly follow from an integration of the data for $\ln\gamma_\pm$ by a procedure we will explain next. That the data of $\phi$ are obtained in this way, makes sense because measuring osmotic pressure this precisely by a direct method would be very difficult, perhaps even impossible. That the procedure below is followed, is supported by the fact that when we use that procedure, we obtain almost exactly the same outcome as was reported, with at most a single point error in the third relevant digit reported (e.g., 0.977 instead of 0.976), and that error can fluctuate in both directions (when it is different, then in some cases the calculated number is higher than reported, and sometimes it is lower). This error does not propagate, i.e., in a list of data at increasing concentration, the single digit error also disappears again. 
In any case, the extremely close match between the calculations based on this procedure, and the reported data, suggests that the procedure below was followed, and the data reported for $\phi$ are indeed based on data for $\ln\gamma_\pm$, and are not independently measured. If that had been the case (data for $\phi$ independently measured), then we would not expect such an extremely close match between the reported data and the outcome of the numerical evaluation that is based on Eq.~\eqref{eq_integration_Hamer_Wu_phi}. 

The procedure we use to generate output for $\phi$, which we are fairly sure must also have been the one used by Hamer and Wu (1972), is to numerically evaluate the integral in Eq.~\eqref{eq_osm_pressure_cell_model_11}, which leads to the iterative procedure to calculate $\phi_i$ according to
\begin{equation}
\phi_i= 1 + \frac{1}{c_{\infty,i}} \sum_i  \frac{\ln \gamma_{\pm,i} - \ln \gamma_{\pm,i-1}  }{\ln c_{\infty,i} - \ln c_{\infty,i-1}} \left( c_{\infty,i} - c_{\infty,i-1} \right) 
\label{eq_integration_Hamer_Wu_phi}
\end{equation}
where $i$ is a counter, running until the last datapoint, but starting only at the second entry, i.e., we start at $i=2$. This is because this procedure does not work for the first entry in the list of $\left(c_{\infty,i},\gamma_{\pm,i}\right)$-data, i.e., when $i \! = \! 1$, then Eq.~\eqref{eq_integration_Hamer_Wu_phi} doesn't work. To obtain a value of $\phi_1$ for the first datapoint, we must assume a certain scaling law to hold in the dilute limit (for concentrations below that first entry), which can either be a square root or cubic root scaling of $\ln\gamma_\pm$ with $c_\infty$. Hamer and Wu used a square root scaling for $\ln\gamma_{\pm}$, which we estimate for their data for NaCl to be $\ln\gamma_\pm = -1.11 \sqrt{c_\infty}$ with $c_\infty$ in mM. Based on this equation we can numerically evaluate the integral in Eq.~\eqref{eq_osm_pressure_cell_model_11} and then we arrive exactly at the first reported number in Table 16 of Hamer and Wu, of $\gamma_\pm \! = \! 0.988$ for $c_\infty \! = \! 1$~mM. 
Interestingly, the exact number calculated for this first entry, does not have much of an influence on the outcome of the integration by Eq.~\eqref{eq_integration_Hamer_Wu_phi} at higher concentrations. For instance, a 10\% change in $\phi$ at 1~mM, has less than 1\% impact on $\phi$ at 10~mM, and no perceptible effect on $\phi$ calculated at $c_\infty \! = \! 50$~mM.

\begin{figure}[!ht]
\centering
\includegraphics[width=0.75\textwidth]{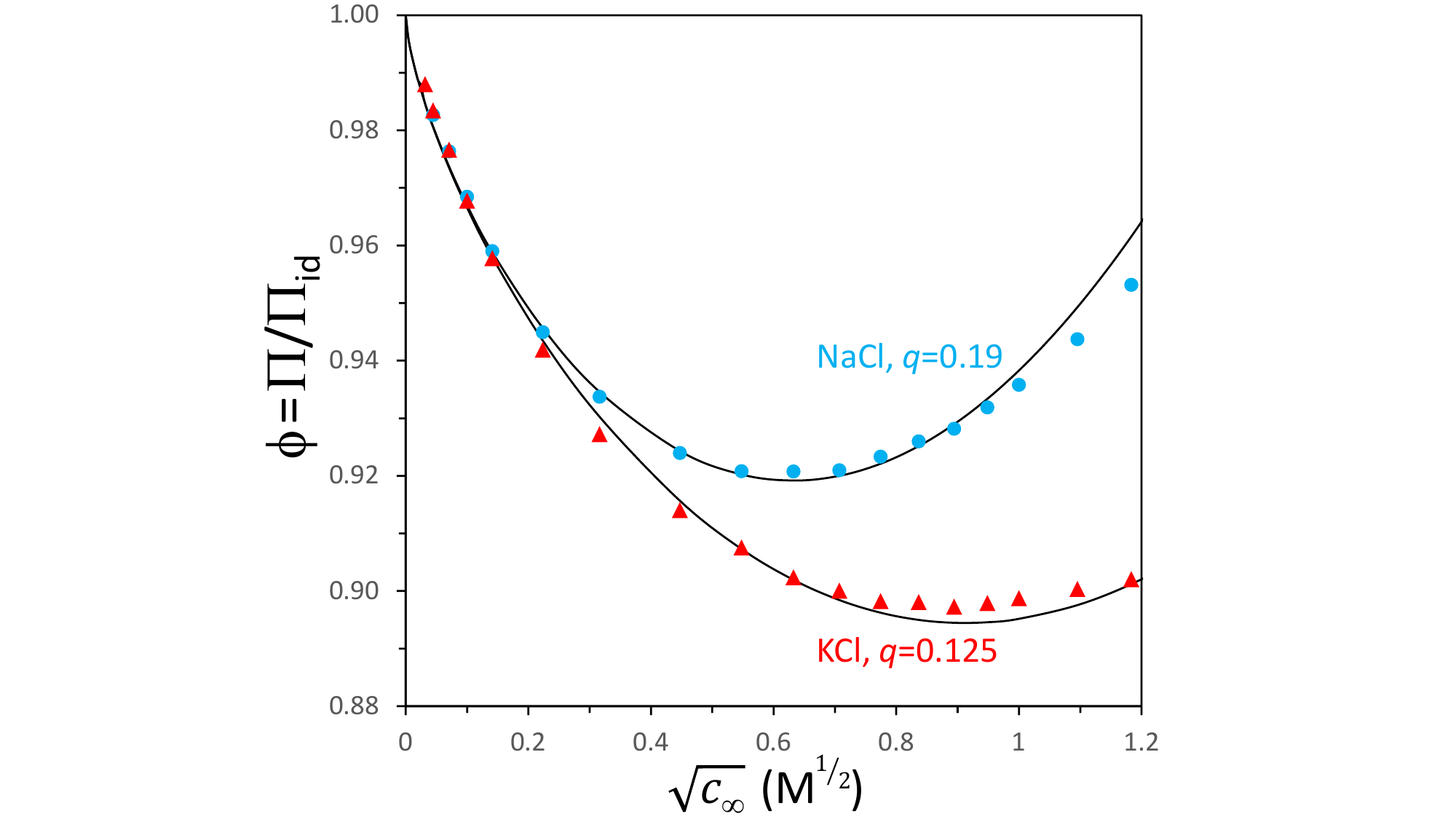}
\vspace{-10pt}
\caption{The osmotic coefficient, $\phi$, of a solution of NaCl or KCl as function of salt concentration, $c_\infty$. The osmotic coefficient is $\phi = \Pi / \Pi\s{id}$, with $\Pi$ osmotic pressure and $\Pi\s{id}=2 c_\infty RT$ the ideal osmotic pressure. Lines are according to Eq.~\eqref{eq_osmotic_pressure_extended} and points are data from Hamer and Wu (1972).}
\label{fig_osm_coeff}
\end{figure}


\end{document}